\documentclass[pdftex,namedreferences]{SolarPhysics}

\usepackage[optionalrh]{spr-sola-addons} 
\usepackage{graphicx}                    
\usepackage{lscape}
\usepackage{longtable}
\usepackage{color}                       
\usepackage{url}                         

\newcommand{\etal}{{\it et al.}}

\newcommand{\aap}{    {\it Astron. Astrophys.}}

\newcommand{\apj}{    {\it Astrophys. J.}}

\newcommand{\nat}{    {\it Nature}}

\newcommand{\pasj}{   {\it Pub. Astron. Soc. Japan}}

\newcommand{\solphys}{{\it Solar Phys.}}

\newcommand{\ssr}{    {\it Space Sci. Rev.}}
\newcommand{\scie}{   {\it Science}}

\begin{document}

\begin{article}

\begin{opening}

\title{Characteristics of EUV coronal jets observed with STEREO/SECCHI}

%
\author{G.~Nistic\`o{}$^{1}$\sep
        V.~Bothmer{}$^{2}$\sep
        S. Patsourakos{}$^{3}$\sep
        G. Zimbardo{}$^{1}$      
       }

%
\runningauthor{Nistic\`o et al.}
\runningtitle{EUV coronal jets}

%
  \institute{$^{1}$ University of Calabria
                     email: \url{nisticogiuseppe@libero.it} email: \url{zimbardo@fis.unical.it}\\ 
             $^{2}$ Goettingen University
                     email: \url{bothmer@astro.physik.uni-goettingen.de} \\
             $^{3}$ Naval Research Laboratory- Also at George Mason University
                     email: \url{spatsourakos@ssd5.nrl.navy.mil} \\
             }

\begin{abstract}
In this paper we present the first comprehensive statistical study of EUV coronal jets observed with the SECCHI (Sun Earth Connection Coronal and Heliospheric Investigation) imaging suites of the two STEREO spacecraft. A catalogue of 79 polar jets is presented, identified from simultaneous EUV and white-light coronagraph observations, taken during the time period March 2007 to April 2008, when solar activity was at minimum. The twin spacecraft angular separation increased during this time interval from 2 to 48 degrees. The appearances of the coronal jets were always correlated with underlying small-scale chromospheric bright points. A basic characterisation of the morphology and identification of the presence of helical structure were established with respect to recently proposed models for their origin and temporal evolution. Though each jet appeared morphologically similar in the coronagraph field of view, in the sense of a narrow collimated outward flow of matter, at the source region in the low corona the jet showed different characteristics which may correspond to different magnetic structures. A classification of the events with respect to previous jet studies shows that amongst the 79 events there were 37 \emph{Eiffel tower}-type jet events commonly interpreted as a small-scale ($\sim35$ arcsec) magnetic bipole reconnecting with the ambient unipolar open coronal magnetic fields at its looptops, and 12 \emph{lambda}-type jet events commonly interpreted as reconnection with the ambient field happening at the bipoles footpoints. Five events were termed \emph{micro-CME} type jet events because they resembled the classical coronal mass ejections (CMEs) but on much smaller scales. The remainig 25 cases could not be uniquely classified. Thirty-one of the total number of events exhibited a helical magnetic field structure, indicative for a torsional motion of the jet around its axis of propagation. A few jets are also found in equatorial coronal holes. In this study we present sample events for each of the jet types using both, STEREO A and STEREO B, perspectives. The typical lifetimes in the SECCHI/EUVI (Extreme UltraViolet Imager) field of view between 1.0 to 1.7 $R_\odot$ and in SECCHI/COR1 field of view between 1.4 to 4 $R_\odot$ are obtained, and the derived speed are roughly estimated. In summary, the observations support the assumption of continuous small-scale reconnection as an intrinsic feature of the solar corona, with its role for the heating of the corona, particle acceleration, structuring and acceleration of the solar wind remaining to be explored in more details in further studies.
             
\end{abstract}

%

\end{opening}

 \section{Introduction}

The twin spacecraft making up the STEREO (Solar TErrestrial RElations Observatory) mission have already provided a rich dataset since launch in October 2006, with images, both in the visible and in the Extreme UltraViolet (EUV) \cite{Kaiser08}. These images are collected by the EUVI (Extreme UltraViolet Imager) A and B imager and COR (CORonagraph) 1 and COR2 A and B coronagraphs on the STEREO A and B satellites as part of the SECCHI (Sun-Earth Connection Coronal and Heliospheric Investigation) instrument packages \cite{Howard08}. Each SECCHI suite consists of the following five telescopes: the EUVI imager and COR1 and COR2 coronagraphs comprise the SCIP package pointing to the Sun, complemented by the off-pointing HI (Heliospheric Imager) 1 and HI2 imagers observing the heliosphere. All imagers together cover the observational range from the Sun to beyond 1 AU.

Besides the classical coronal mass ejections (CMEs) that are identified in the FOVs of the COR imagers ({\it e.g.} \opencite{HowardT08}), small scale collimated ejections of solar plasma are another common feature ({\it e.g.} \opencite{Patsourakos08}).
These coronal jets are best observed inside polar coronal holes at EUV wavelengths when the plasma beams are seen in emission against the dark background and are not obscured by bright ambient coronal structures.  
Very early observations of coronal jets were provided by Skylab and, in more details, later by Yohkoh in the soft X-rays \cite{Shimojo96,ShimojoS00}. Polar coronal jets were studied by \inlinecite{Wang98} using images of the LASCO (Large Angle and Spectrometric COronagraph) and EIT (Extreme ultraviolet Imaging Telescope) instruments on board SOHO (Solar and Heliospheric Observatory).
Further observations were reported by \inlinecite{Alexander99} with the TRACE (Transition Region And Coronal Explorer) spacecraft.
More recently, also Hinode has provided important data on polar jet parameters \cite{Savcheva07,Kamio07,Moreno-Insertis08,Filippov09}. X-ray jets have typical lengths of $10^4$--$4\times 10^5$ km, widths of $5\times 10^3$--$ 10^5$ km, and speeds ranging from 10 to 1000 km/s \cite{Shimojo96}. It is usually assumed that the jet is the result of magnetic reconnection phenomena happening in the solar corona \cite{Shibata92, Yokoyama95, Yokoyama96, Moreno-Insertis08, Pariat09}. Coronal jets can be observed both in active regions, quiet sun, and coronal holes \cite{Shimojo96}. 
A polar coronal jet has been  studied for the first time stereoscopically with observations from the STEREO/SECCHI imagers by \inlinecite{Patsourakos08}, demostrating the existence of helical magnetic field lines in this event.

The two points of view provided by the twin STEREO satellites are very helpful for the identification of coronal hole jets. Indeed, for the first time it is possible with STEREO/SECCHI data to assess what is the three dimensional (3D) structure of the jet, to understand what projection effects are present in single point observations, and to try to establish the true 3D velocity for the jet \cite{Patsourakos08}.

Here we present the first catalogue of STEREO polar coronal jets, comprising 79 events, which were observed by both EUVI imagers in the ultraviolet, and contemporarily by the COR1 coronagraph in white-light, during the period at solar minimum at the end of solar cycle 23 from March 2007 to April 2008. This time interval, during which the separation angle between the two spacecraft increased from 2 to 48 degrees with respect to the Sun, includes two time periods of high time cadence observations in May 2007 and January 2008. 

The criteria according to which coronal jet events were identified and selected from the SECCHI dataset are explained in Section 2 of this paper, including some brief summary of their basic characteristics. Prototype events which also emphasize the difference of the viewpoints when seen either from STEREO A or from STEREO B, are shown in Section 3. In Section 4 the lifetime distributions of jets in the different wavelengths in the EUVI and in COR1 FOV are discussed. The conclusions are given in Section 5, followed by the catalogue of polar coronal jets in the Appendix. 
\section{Selection criteria and statistics of the jet catalogue}

SECCHI images are taken at four EUV wavelengths centered at 171 \AA, 195 \AA, 284 \AA~with a time cadence of 2.5, 10, 20 minutes, and at 304 \AA~with a time cadence of 10 minutes. During high time cadence observations on May 5--19, 2007, and on January 7--20, 2008 the time cadences was 75 s in the 171 channel. The pixel size of the EUVI CCD on the solar disk has a spatial resolution of 1.6 arcsec. Because of the low time cadence of the 284 \AA~observations, in this jet study more events are seen in the 171, 195 and 304 \AA~wavelength ranges. We began our systematic study of coronal jets in the SECCHI EUVI A and B data after STEREO's completed in-orbit insertion and the official science mission start on January 22, 2007, and after some follow on period of routine imaging, in March 2007. 
For the search of coronal jets EUVI images were used that are provided at the SECCHI website \url{http://secchi.nrl.navy.mil/}. SECCHI EUVI A and B daily movies are available at \url{http://cdaw.gsfc.nasa.gov/stereo/daily_movies/}. The existence of subsequent jets in the COR1 A and B FOVs were studied in the daily movies in intensity running difference available at \url{http://cor1.gsfc.nasa.gov/dailymov/MPEG/}. The running difference movies allow to identify faint transient coronal structures, {\it e.g.} those hidden by coronal streamers. 

\begin{figure}[htbp]
     \begin{center}
       \begin{tabular}[htpb]{l l}
        \includegraphics[width=5.6 cm]{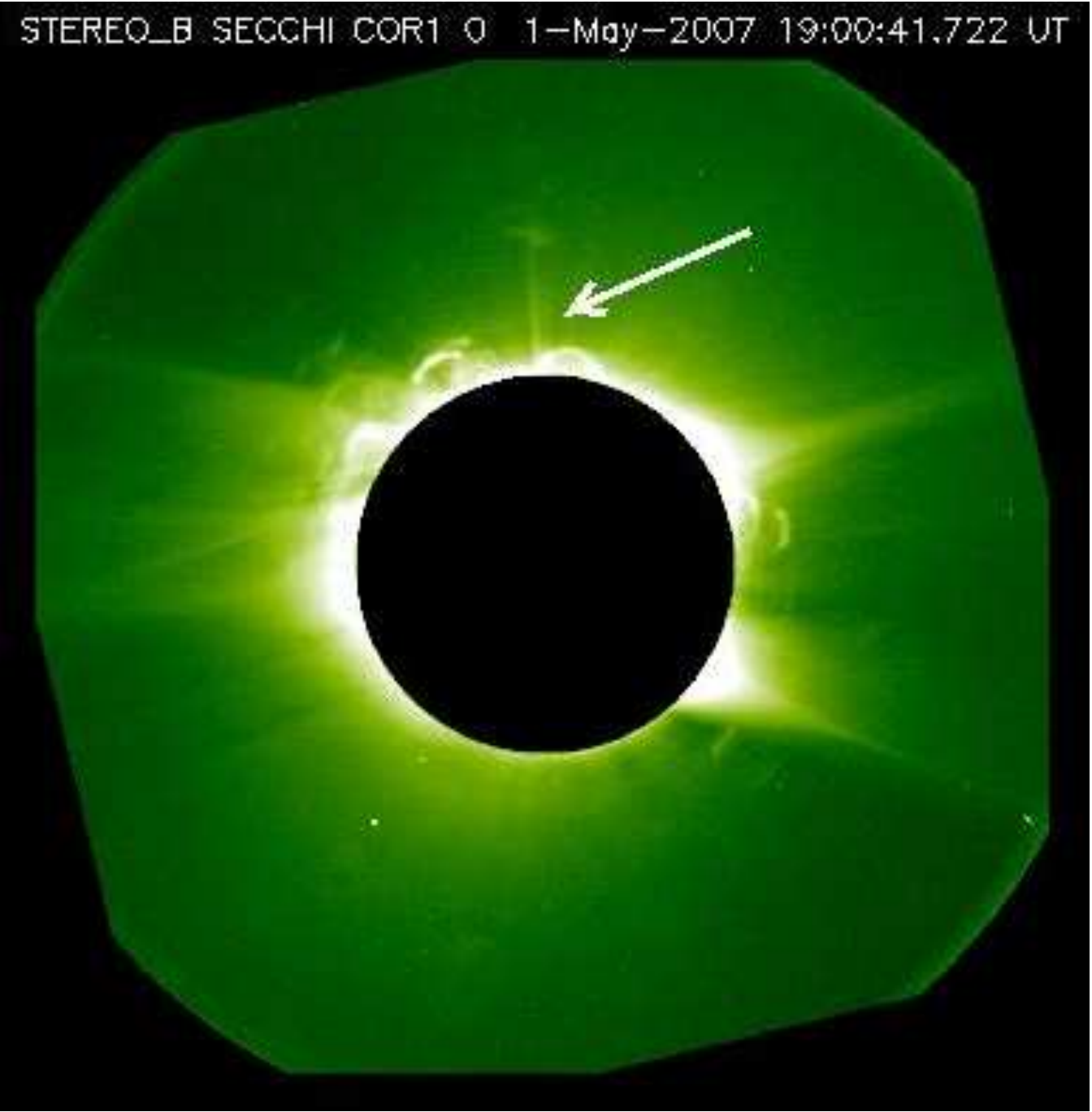} &
        \includegraphics[width=5.6 cm]{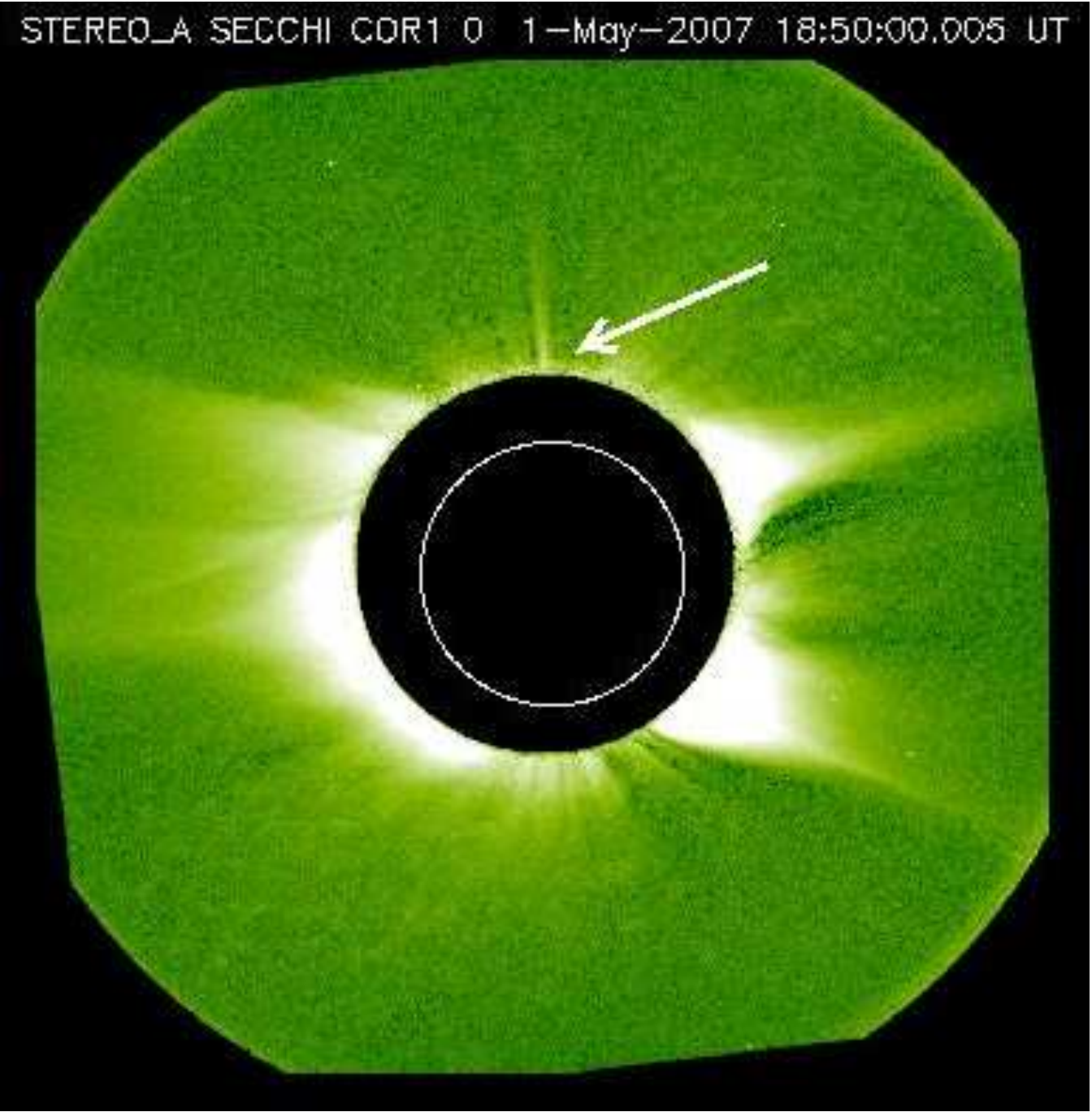}\\
      \end{tabular}
     \end{center}
     \caption{Coronal jet seen in the COR1 field of view for STEREO B (left) and A (right) on 01 May 2007 near the Sun's north pole (event no. 10 in the catalogue). Note the different offset of the  STEREO A and B COR1 occulters with respect to the Sun's center due to different pointing. The spacecraft angular separation was $\Delta \phi_{AB} \simeq 6.18^{\circ}$.}
   \label{fig_1}
  \end{figure}

In order to carry out a quantitative study of coronal jets, after identification of the events, SECCHI data have been analysed with the \emph{SolarSoft} package (\url{http://www.lmsal.com/solarsoft/}), which provides routines for sophisticated SECCHI image processing. The data are processed with the \texttt{secchi\_prep} routine and specific commands such as (\texttt{plot\_map, scc\_triangulate, cursor})  are used in order to obtain the values for the position angle of the events at the solar limb.

A selection criterion applied to the list of candidate jet events derived from the SECCHI images and COR difference movies, covering 256 events, was the unique visibility of the jets not just in the 171 \AA~ or 195 \AA~observations of EUVI but also at least in either the COR1 A or COR1 B observations. This criterion ensures that a jet is not confined to the low coronal FOV of EUVI, {\it i.e.} to a height of 1.7 $R_{\odot}$, and does not simply be a spicule or macrospicule event \cite{Wang98, Yamauchi04}. The visibility of the jet in the 171 \AA~ or the 195 \AA~ lines implies that the observed jet is comprising hot plasma at typical temperatures of $\sim10^6$ K. The condition that jets be detected by both STEREO A and STEREO B EUV imagers reduces the number of events to 79. 
\begin{table}[htpb]
\begin{tabular}{l c c c c} 
\hline
\bfseries{Classification} &\bfseries{Number of events} &\bfseries{Limb}& \bfseries{Edge} & \bfseries{Inside}\\
\hline
Eiffel Tower   &  19/37 &  21 &  2 & 14 \\
Lambda         &   4/12 &   4 &  5 & 3  \\
Micro-CME       &   2/5  &   4 &  1 &    \\
Not classified &   6/25 &  17 &  4 & 4  \\
\hline
Helical structure &  31 & & & \\
\hline 
\end{tabular} 
\caption{Statistics of jet morphology. In the Table we give the event-type, the number of helical events over the total number for the specific type, the locations in which they seem to occur. Edge (Inside) indicates jets which are close to (detached from) the coronal hole boundary. In the last row the total number of events that show helical features.}
\label{tab_1}
\end{table}
Fig. \ref{fig_1} shows an example of a typical coronal jet identified in the COR1 B and A fields of view.

 In the catalogue in the Appendix we provide information on the date for all of the identified 79 jet events, the angular separation $\Delta \phi_{AB}$ between STEREO A and STEREO B as seen from the Sun at the time of the events, the time of observation with EUVI at the different wavelengths and with COR1, the position angle $\alpha$ and $\beta$, respectively in the EUVI and COR1 imagers at the solar limb, measured positive counterclockwise from solar North and, when possible, the result of the assessment of the jet's morphology. 
The spatial distribution of jet positions corresponds to 45 events (57\%) found in north polar coronal holes (NPCHs) and to 34 events (43\%) in south polar coronal holes (SPCHs). The question whether the slight difference in the number of northern and southern events is related to a different areal size of the coronal holes in both solar hemispheres is beyond the scope of the present study.
\begin{figure} 
\centerline{\includegraphics[width=0.9\textwidth,clip=]{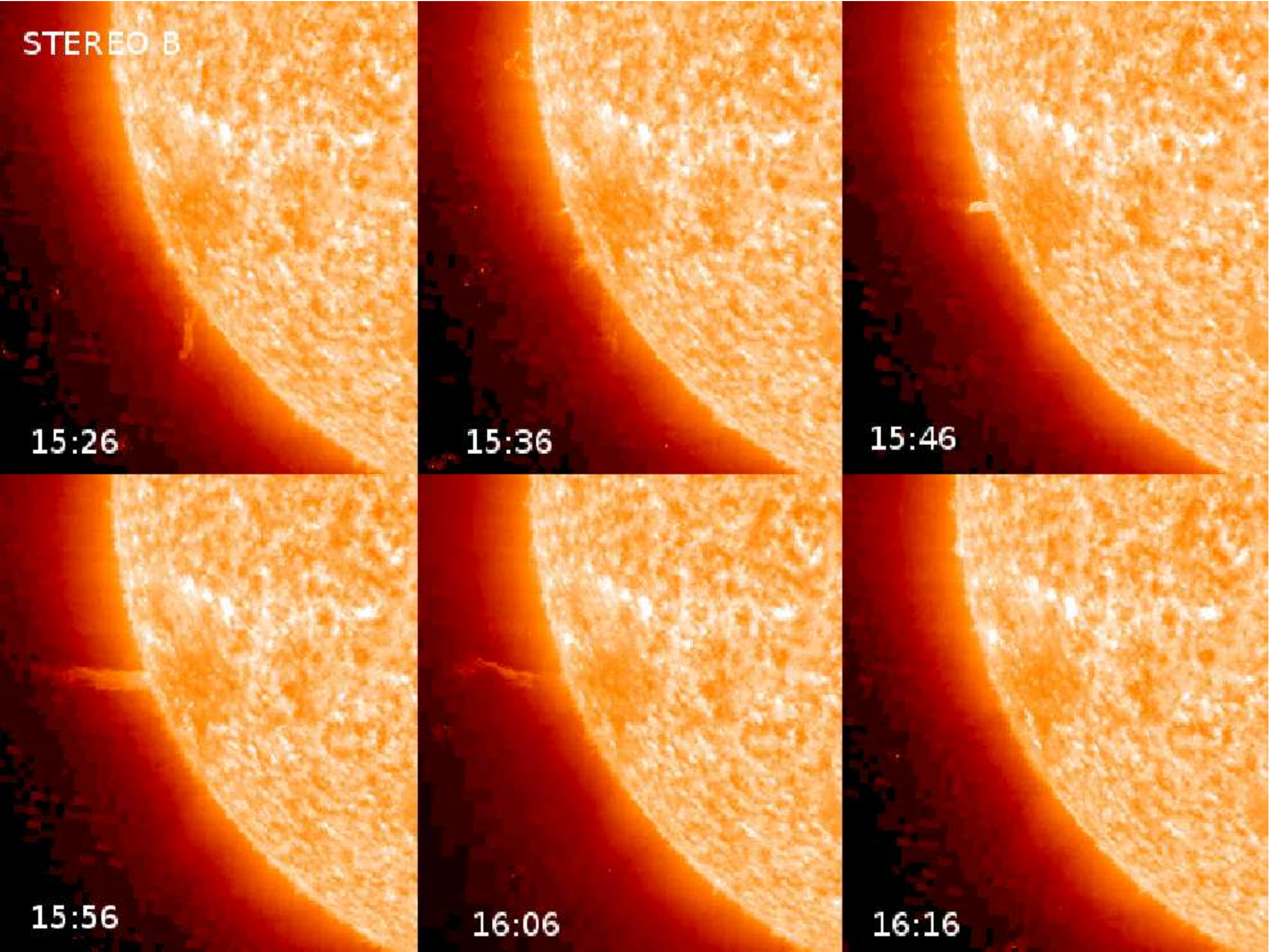}}
 \caption{Sequence of images from STEREO B showing the evolution of an equatorial coronal jet on November 11, 2007.}
\label{fig_2}
\end{figure}
A number of equatorial jets have also been found in the course of the search for coronal jet events, indicating that coronal jets are an overall coronal feature rather than limited to the polar regions, in agreement with the findings of \inlinecite{Moreno-Insertis08}. Fig. \ref{fig_2} shows a sample event as a sequence of images from STEREO B at 304 \AA, in which the equatorial jet on November 11, 2007 has been identified. The corresponding equatorial coronal hole can be seen as a darker cooler area. 
The event was also visible in the 171 \AA~ and 195 \AA~ wavelengths.
   
However, because of the usual presence of large-scale equatorial coronal structures, as is typical near solar minimum, {\it e.g.} the presence of a streamer belt, only for very few equatorial jet events a clear subsequent signal could be identified in the COR1 data without improved image analysis. The equatorial jets identified so far from the SECCHI data are not included in the jet catalogue presented in the Appendix, but will be the subject of further analysis. 
\section{Typical morphology of coronal jets}

\begin{figure}[htbp]
     \begin{center}
       \begin{tabular}[htpb]{l l}
        \includegraphics[width=5.6 cm]{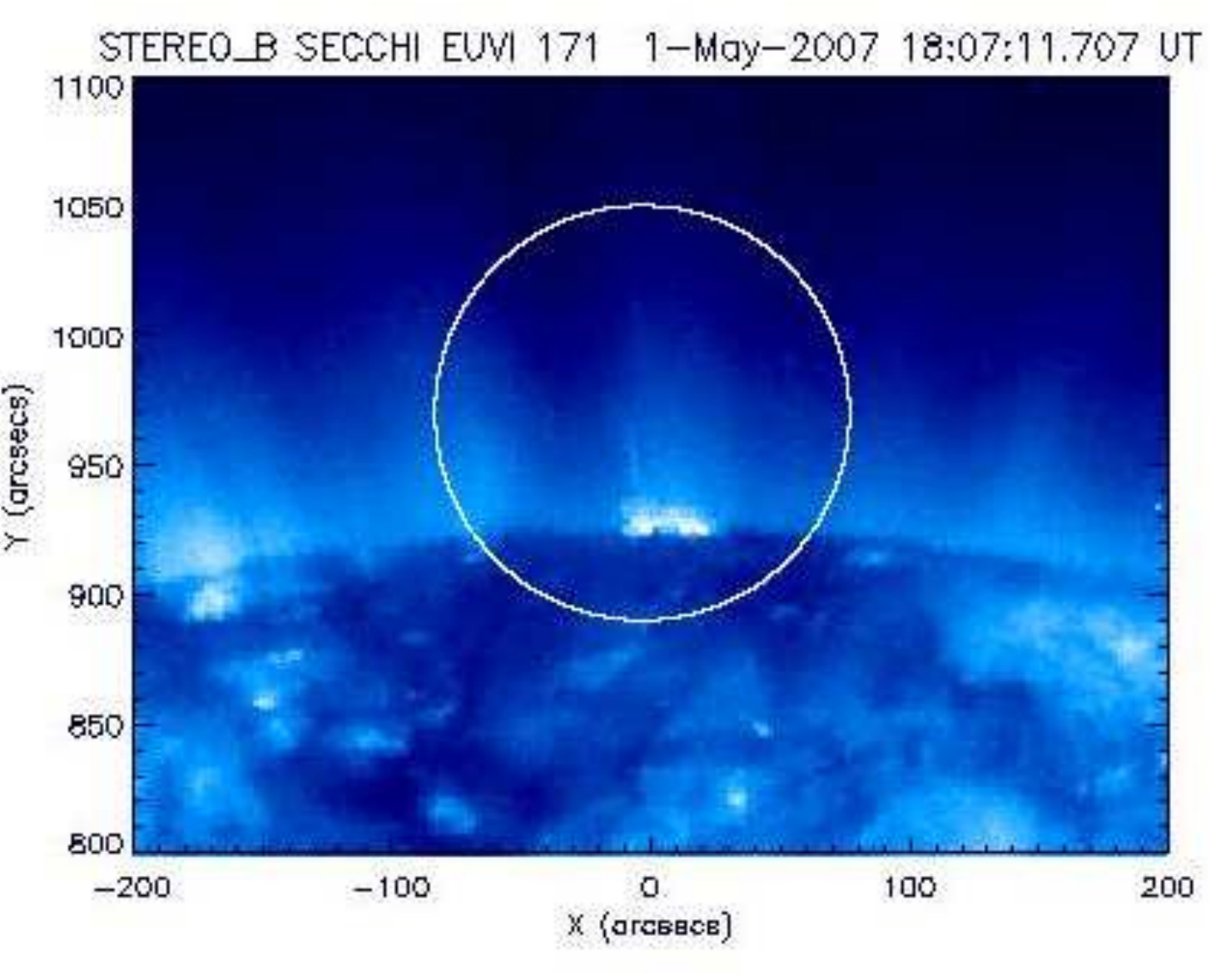} &
        \includegraphics[width=5.6 cm]{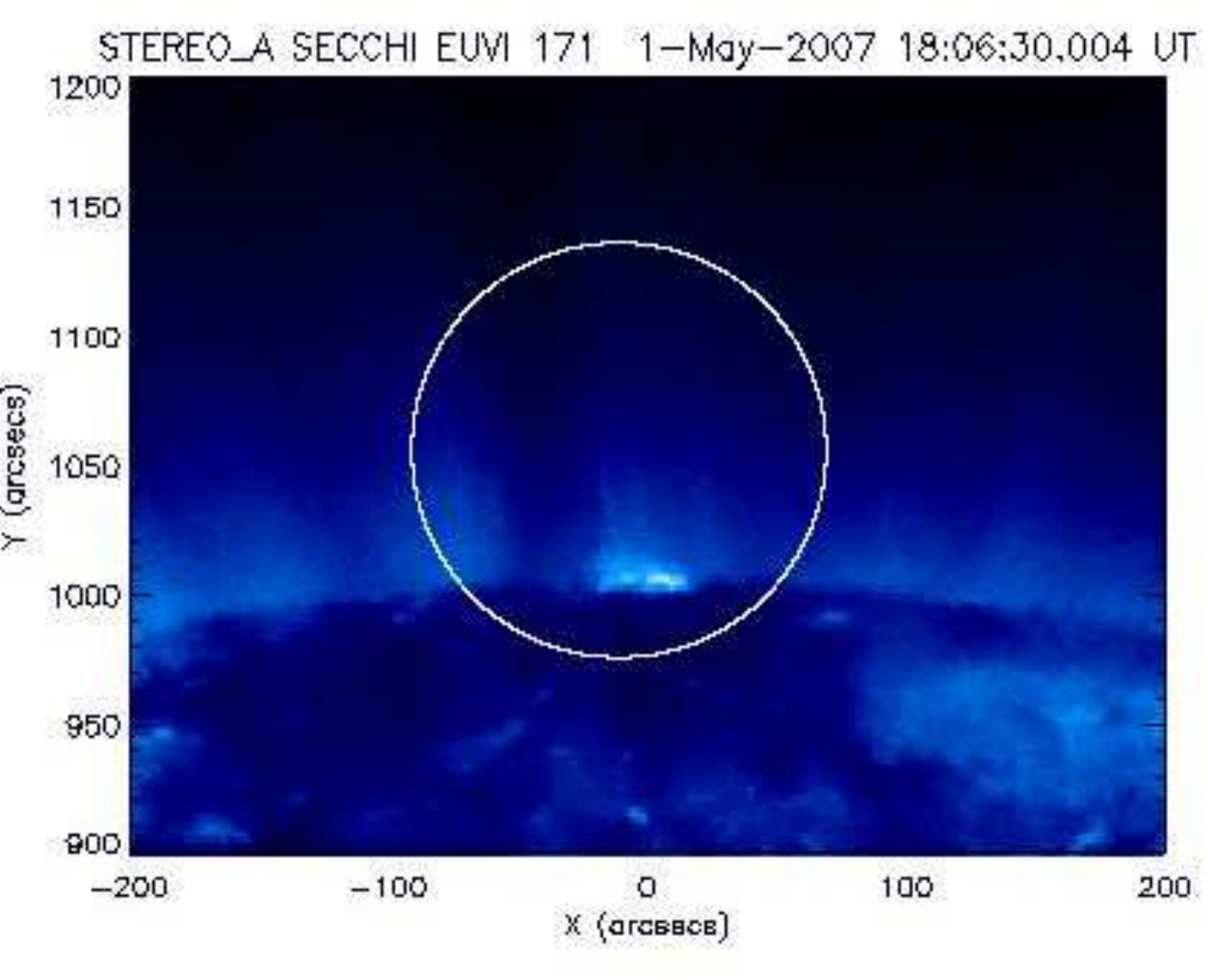}\\
        \includegraphics[width=5.6 cm]{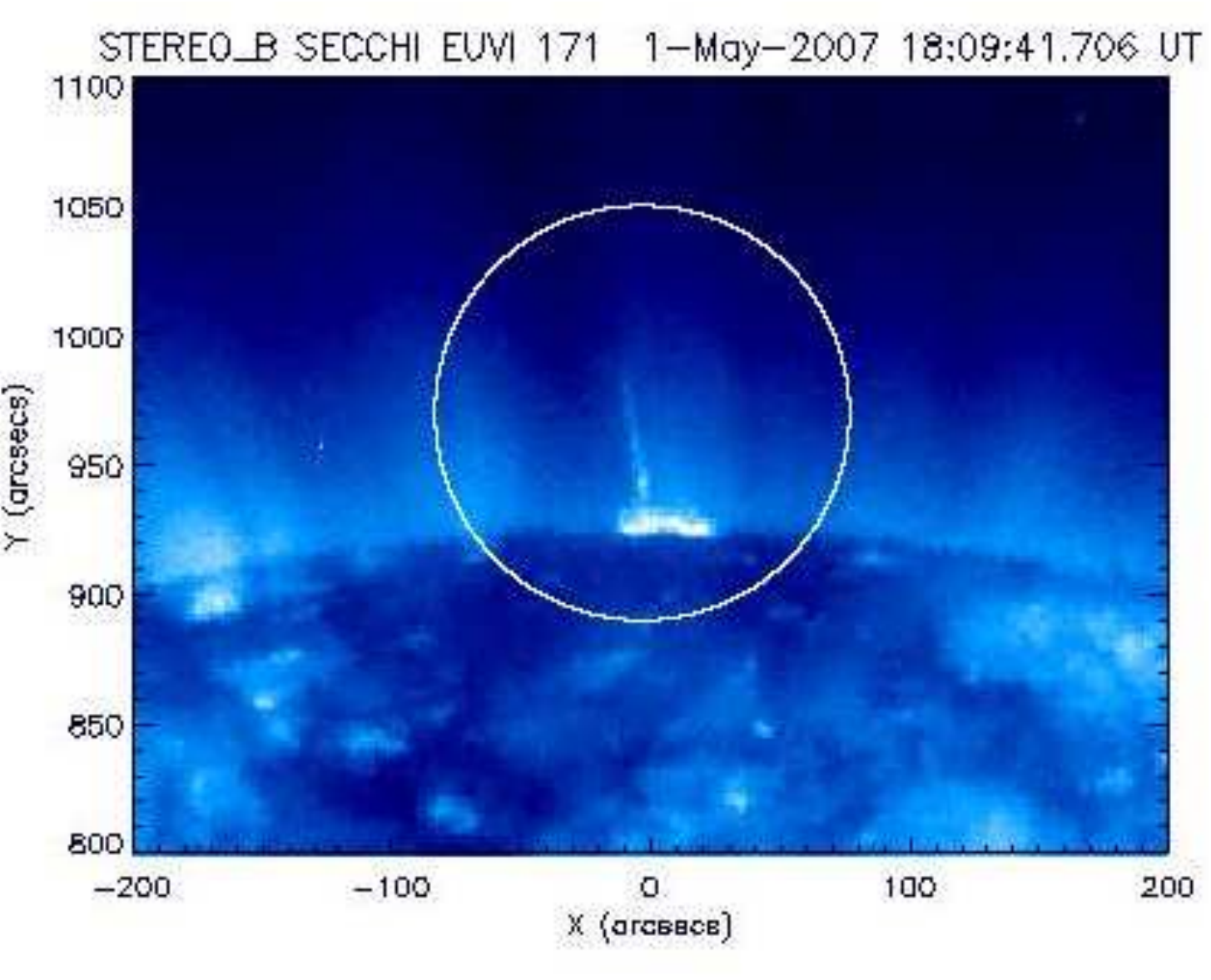} &
        \includegraphics[width=5.6 cm]{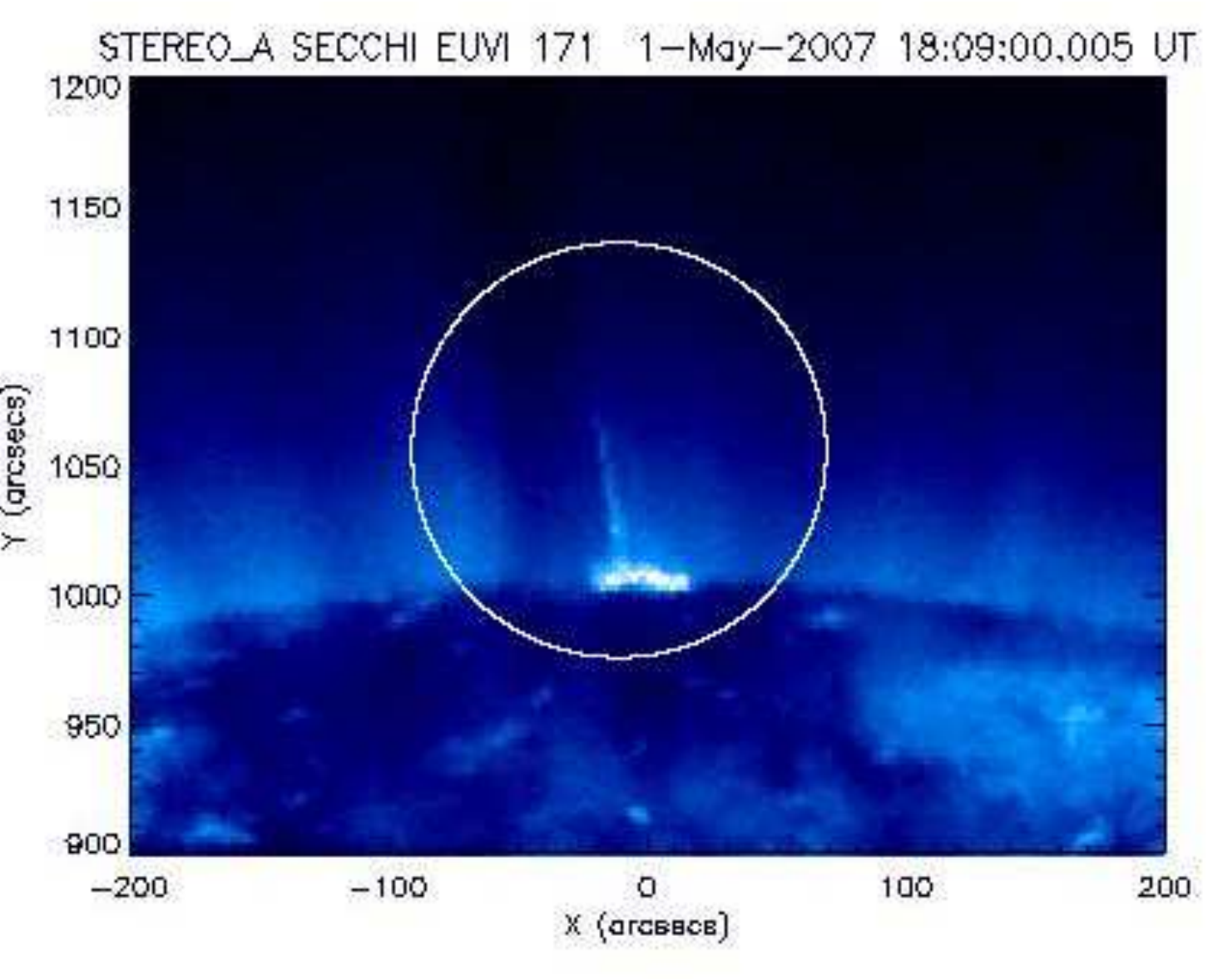}\\
        \includegraphics[width=5.6 cm]{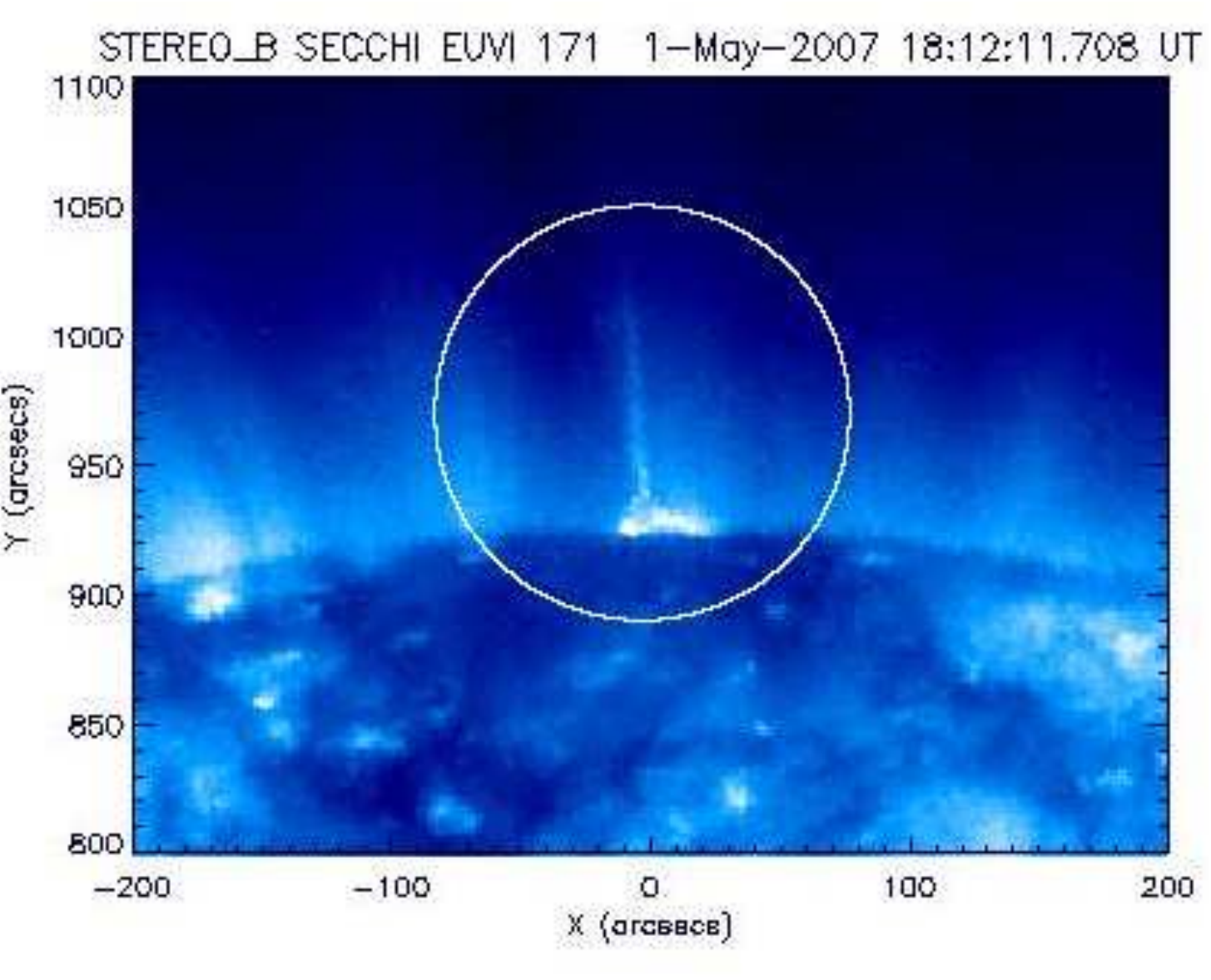} &
        \includegraphics[width=5.6 cm]{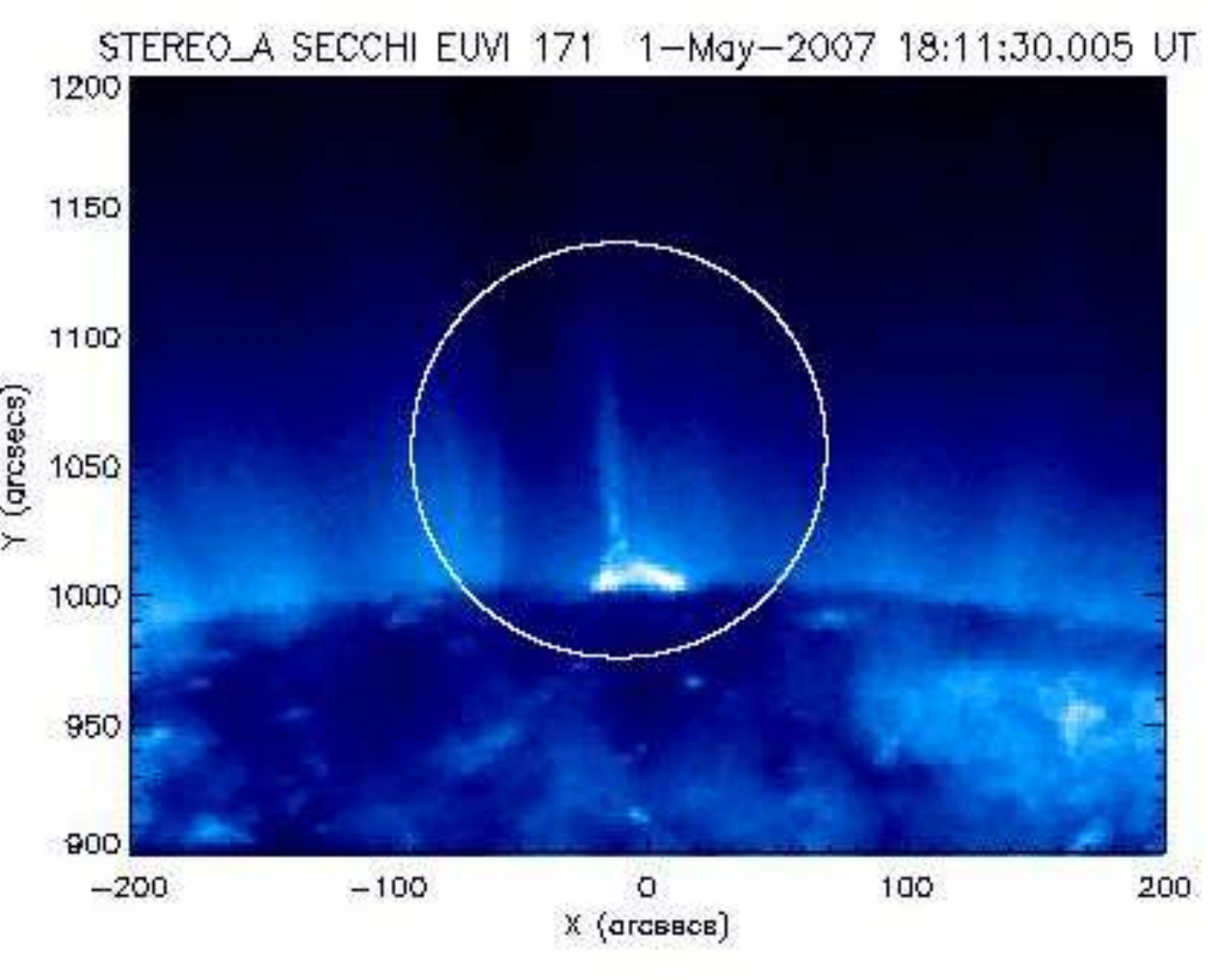}\\

        \end{tabular}
     \end{center}
     \caption{Sample ``\emph{Eiffel Tower}'' event in the north polar coronal hole on 01 May 2007 imaged at 171 \AA. Left: view from STEREO/SECCHI EUVI B. Right: view from STEREO/SECCHI EUVI A. Note that in this case the small loop below the jet is clearly resolved. Same events as in Fig. 1.}
   \label{fig_3}
  \end{figure}

\begin{figure}[htbp]
     \begin{center}
       \begin{tabular}[htpb]{l l}
        \includegraphics[width=5.6 cm]{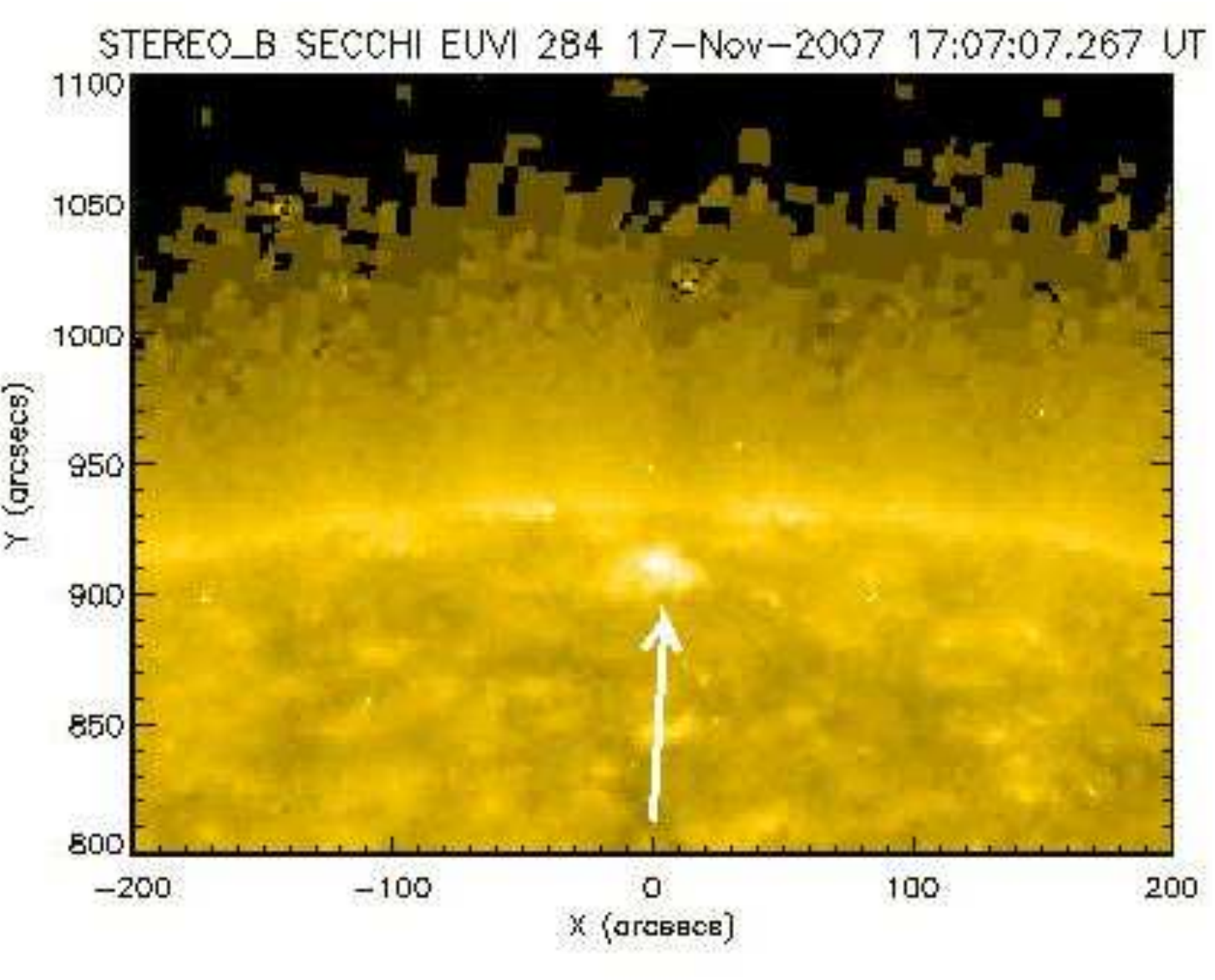} &
        \includegraphics[width=5.6 cm]{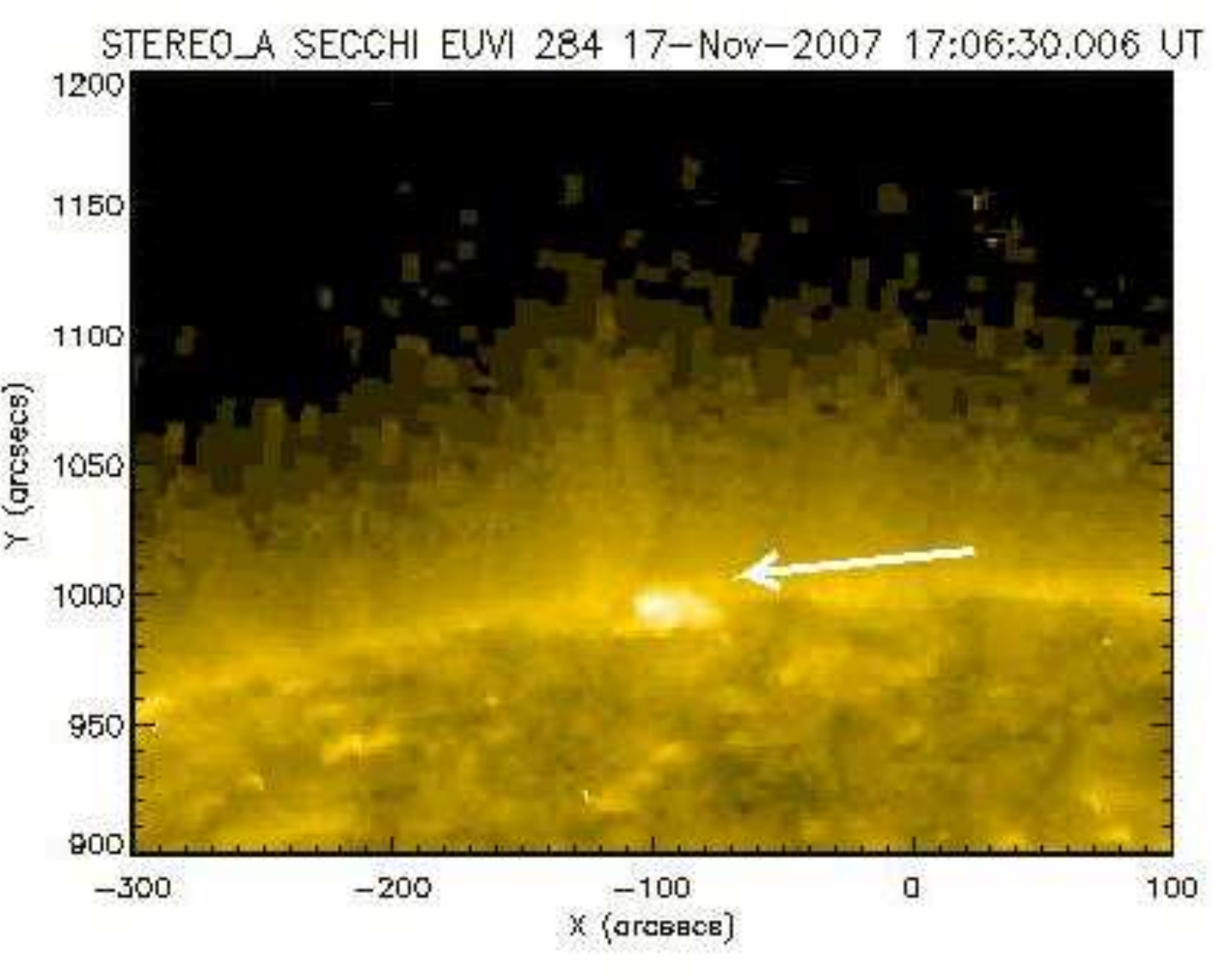}\\
        \includegraphics[width=5.6 cm]{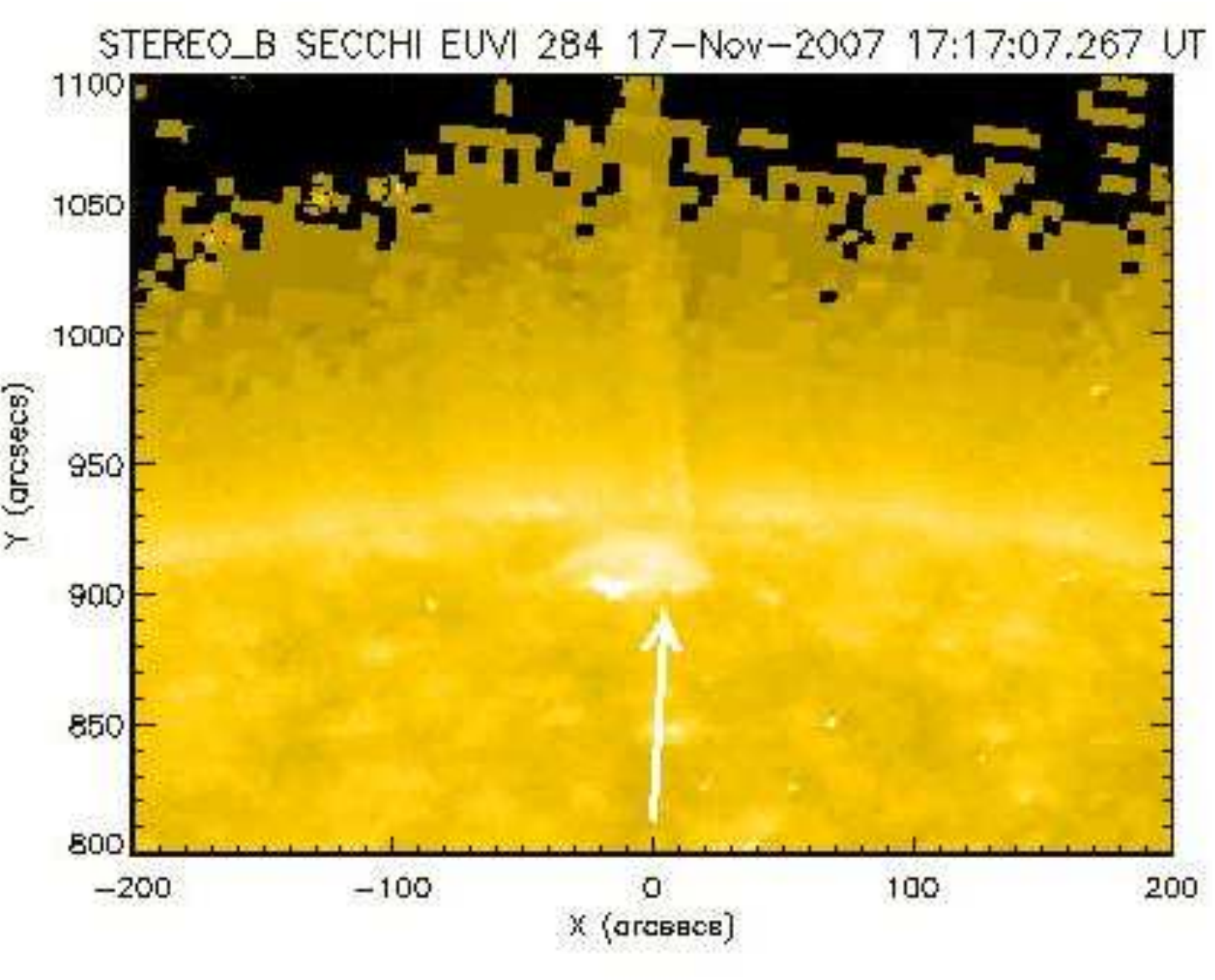} &
        \includegraphics[width=5.6 cm]{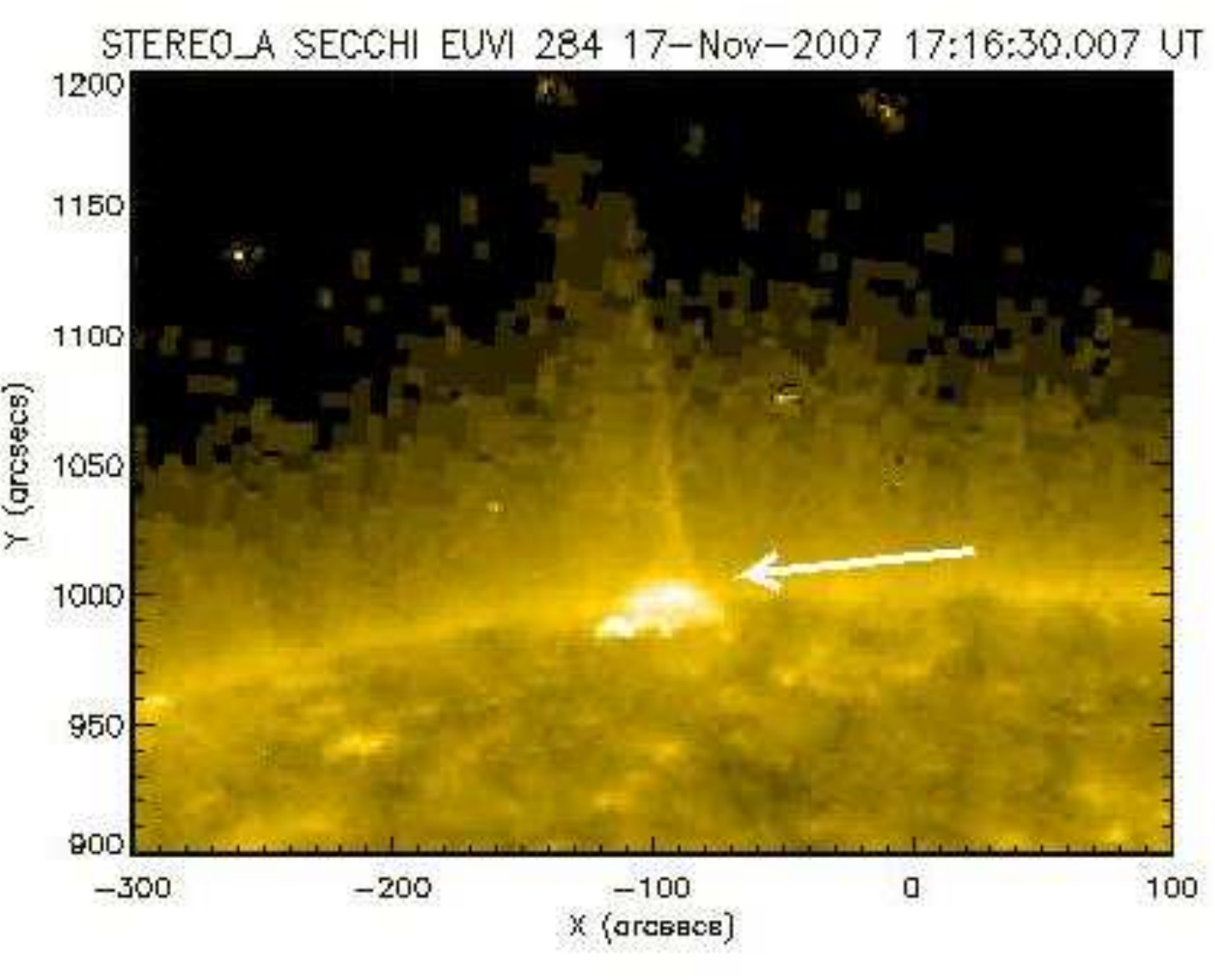}\\
        \includegraphics[width=5.6 cm]{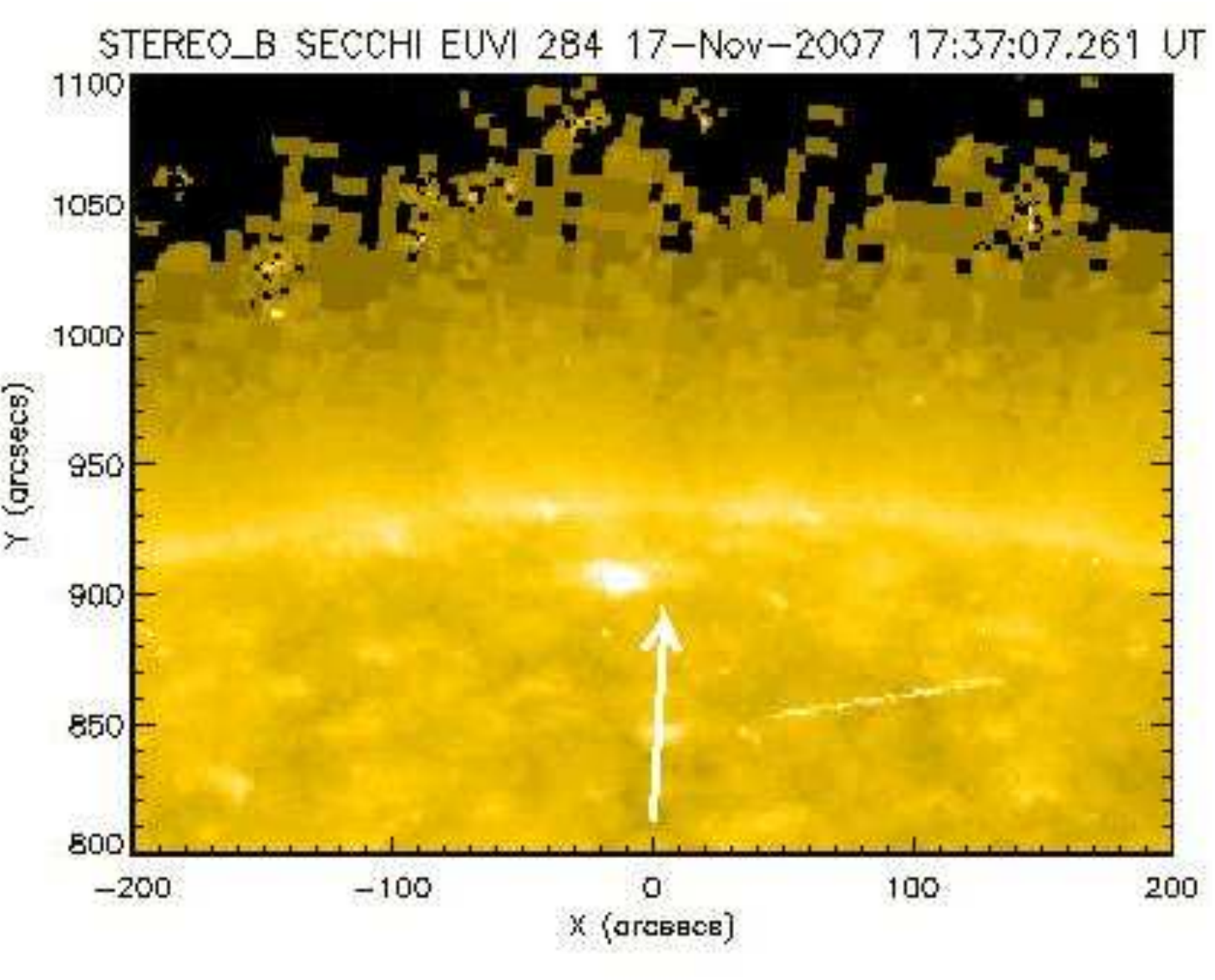} &
        \includegraphics[width=5.6 cm]{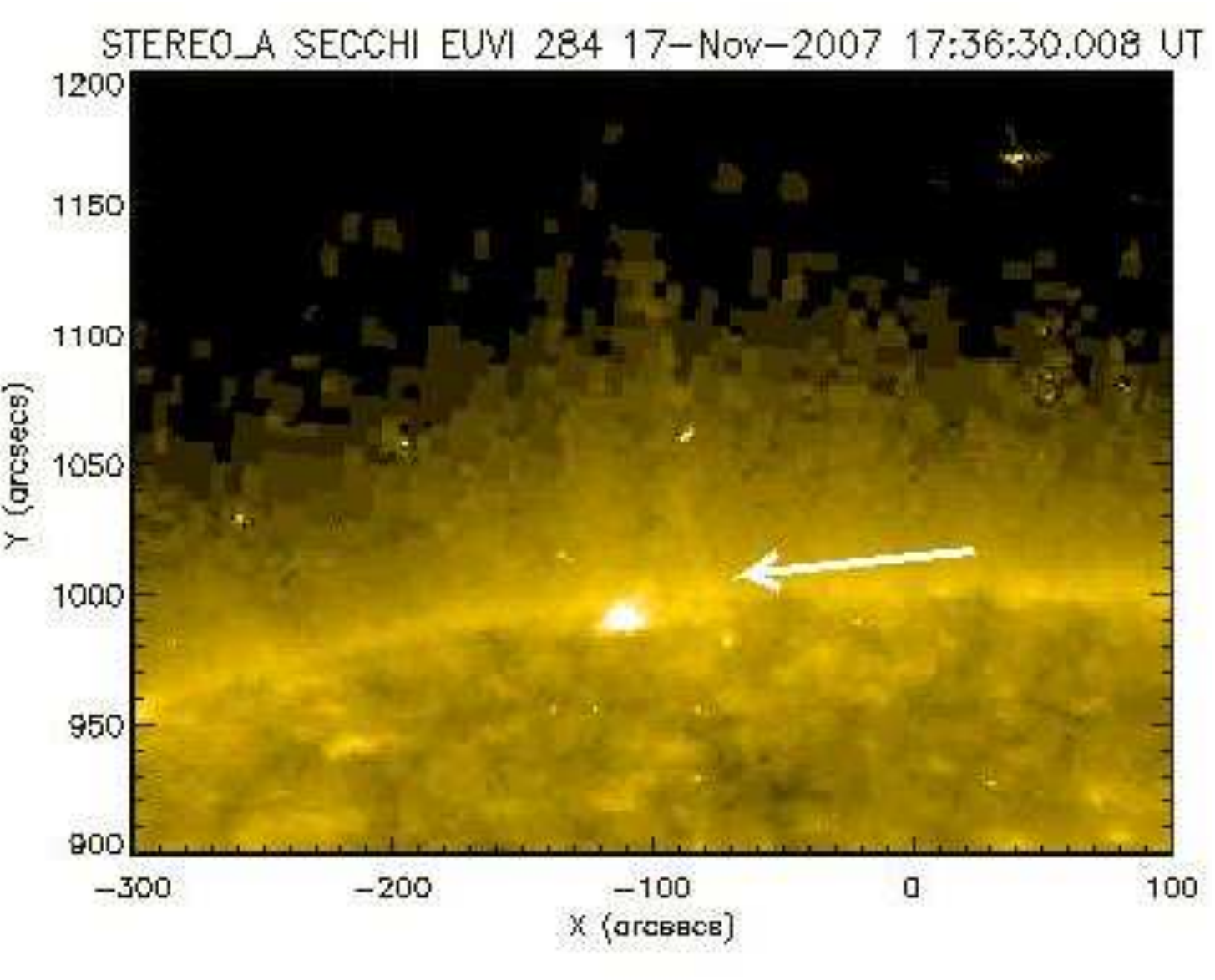}\\

        \end{tabular}
     \end{center}
     \caption{Sample ``$\lambda$'' event in the north polar coronal hole on 17 November 2007 as observed by SECCHI EUVI A and B at 284 \AA. Spacecraft angular separation: $\Delta \phi_{AB} \simeq 40.57^{\circ}$. Left: view from STEREO B. Right: view from STEREO A.}
   \label{fig_4}
  \end{figure}

Based on previous results coronal jets can be basically classified into the following two categories (Table \ref{tab_1}):
i) \emph{Eiffel Tower} (ET) jets which resemble a shape like small helmet streamers or an inverted-Y shape, and may correspond to the magnetic topology of a small-scale photospheric bipole reconnecting with ambient open unipolar field lines of opposite polarity at its looptops \cite{Shimojo96, Yamauchi04, Patsourakos08, Pariat09,Filippov09}; 
ii) \emph{$\lambda$-jets} in which a small-scale photospheric magnetic bipole reconnects with ambient unipolar field lines near its footpoints \cite{Shibata92, Yokoyama96,Filippov09}.
It should be noted that the EUVI observations only allow us to distinguish whether a jet occurred close to loop tops or the loop footpoints, {\it i.e.} we do not observe directly the magnetic field structure. An ideal case would thus be the phase of the STEREO mission when the viewing angle with respect to the Sun-Earth line had increased to about 90 degrees to correlate limb observations from STEREO with disk centered magnetograms from Hinode and SOHO.
The morphology for each of the jet events listed in the Table of the Appendix was investigated for its morphological characteristics according to the above two categories. Fig. \ref{fig_3} and Fig. \ref{fig_4} show archetype ``Eiffel tower'' and ``lambda'' events observed by SECCHI EUVI A and B.
Fig. \ref{fig_3} shows a sequence of images of a north polar coronal jet taken at 171 \AA, seen by both STEREO A  and STEREO B on May 1, 2007 (event no. 10 in the catalogue). The fast ejection of hot material, as well as the bright loop at the bottom of the jet, are clearly seen by both spacecraft. The angular separation was $\Delta \phi_{AB} \simeq 6.18^{\circ}$, and the difference in the viewpoints is already becoming obvious and it is also possible to identify the helical structure of the ejection.

\begin{figure} 
       \begin{tabular}[htpb]{l l}
        \includegraphics[width=5.6 cm]{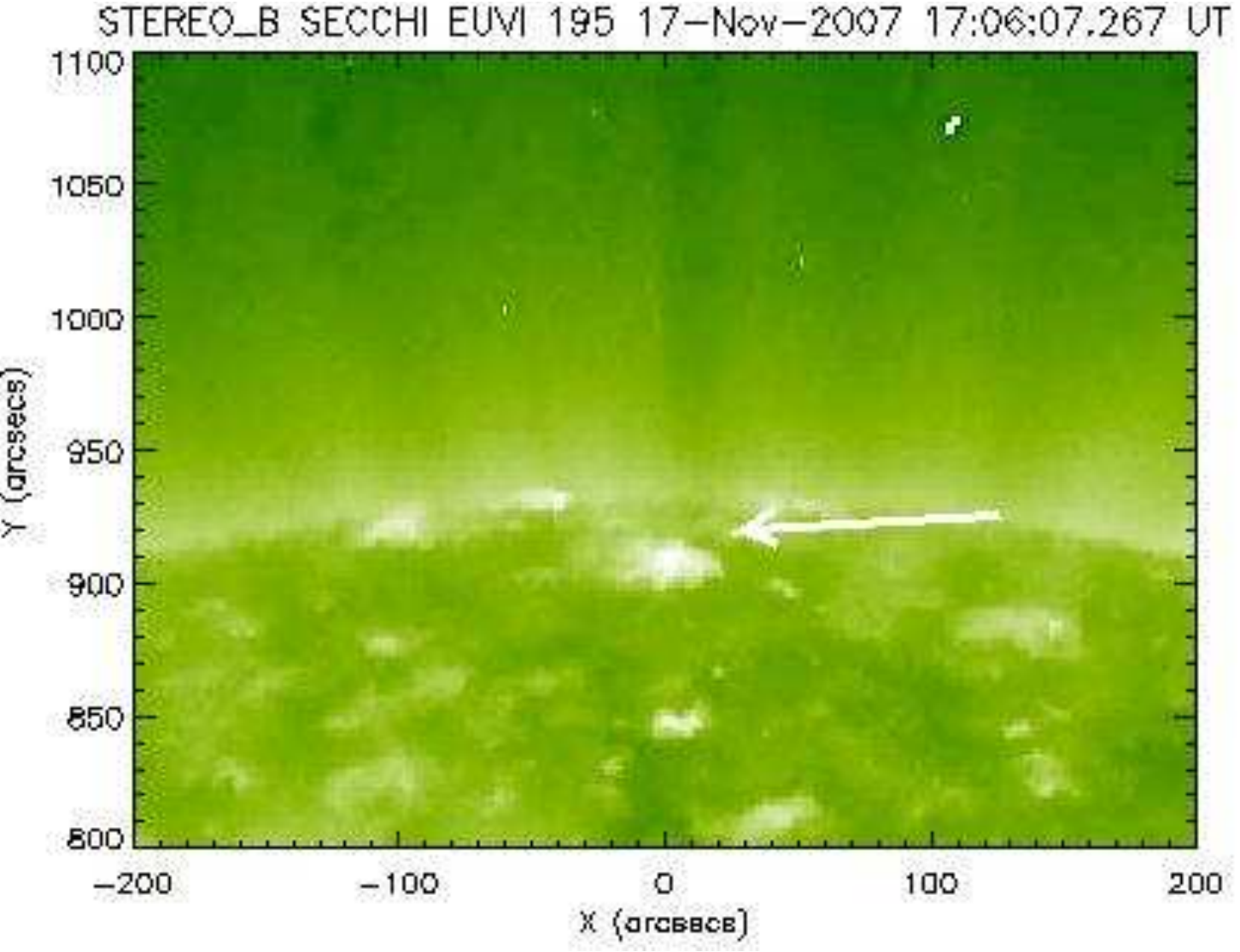} &
        \includegraphics[width=5.6 cm]{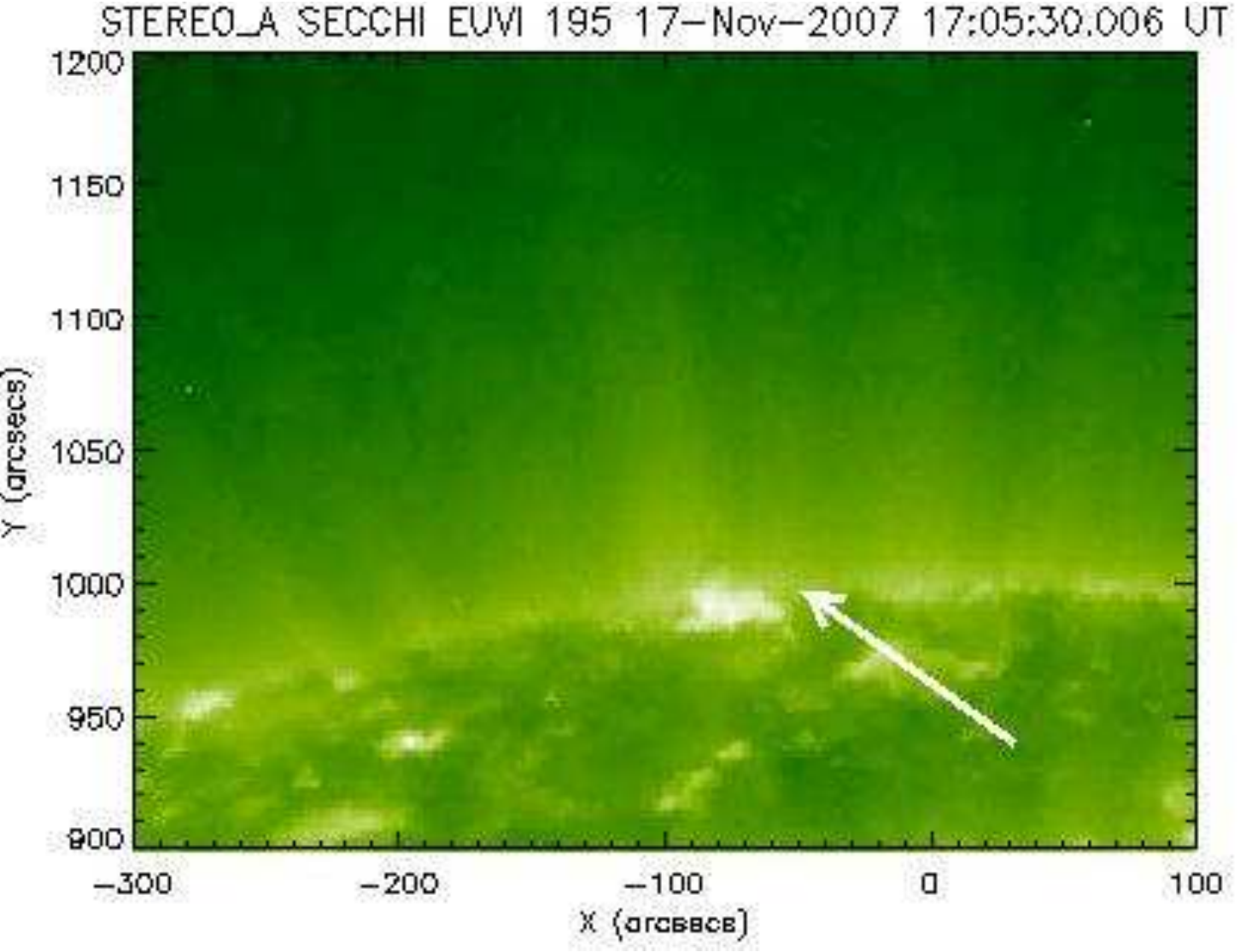}\\
        \includegraphics[width=5.6 cm]{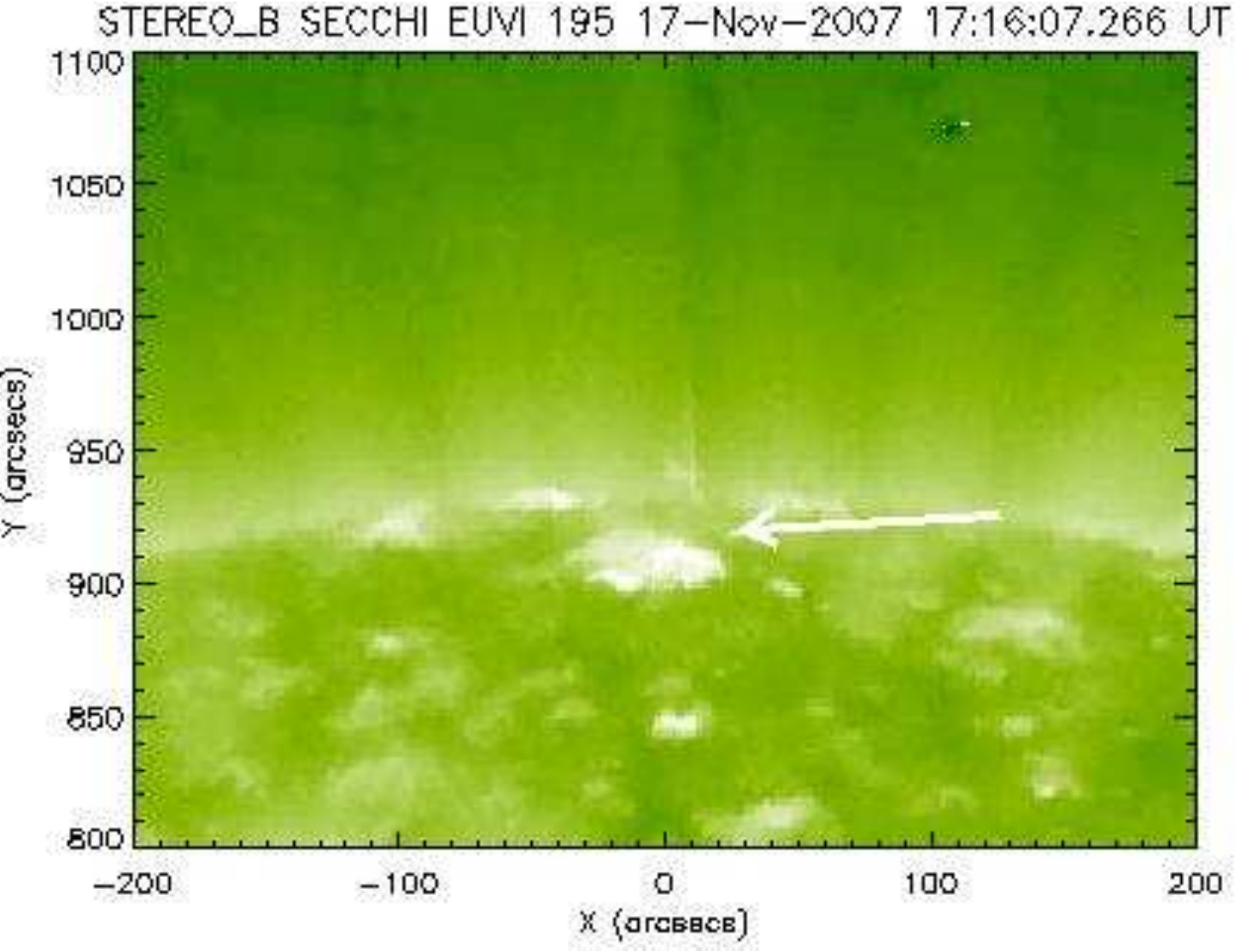} &
        \includegraphics[width=5.6 cm]{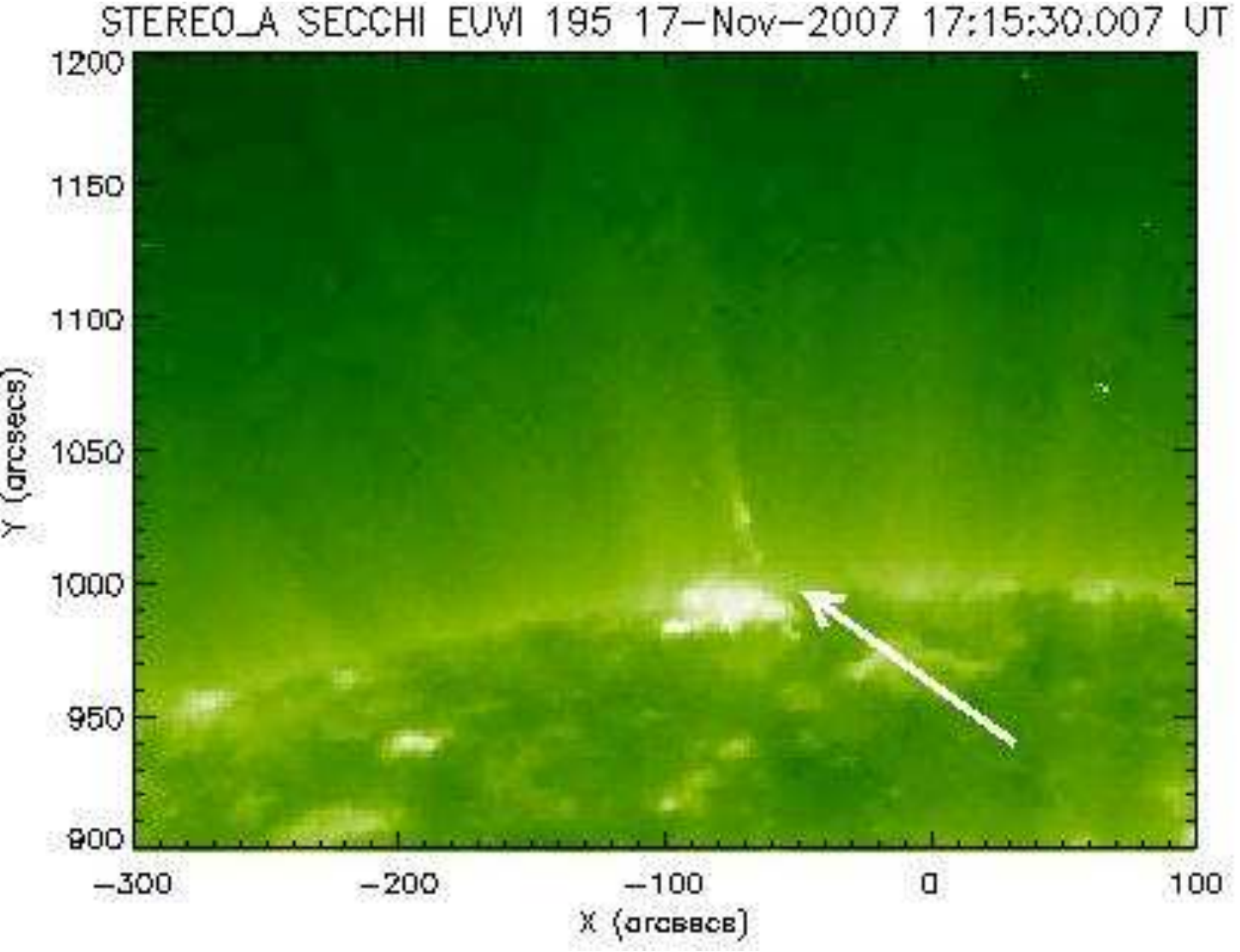}\\
        \includegraphics[width=5.6 cm]{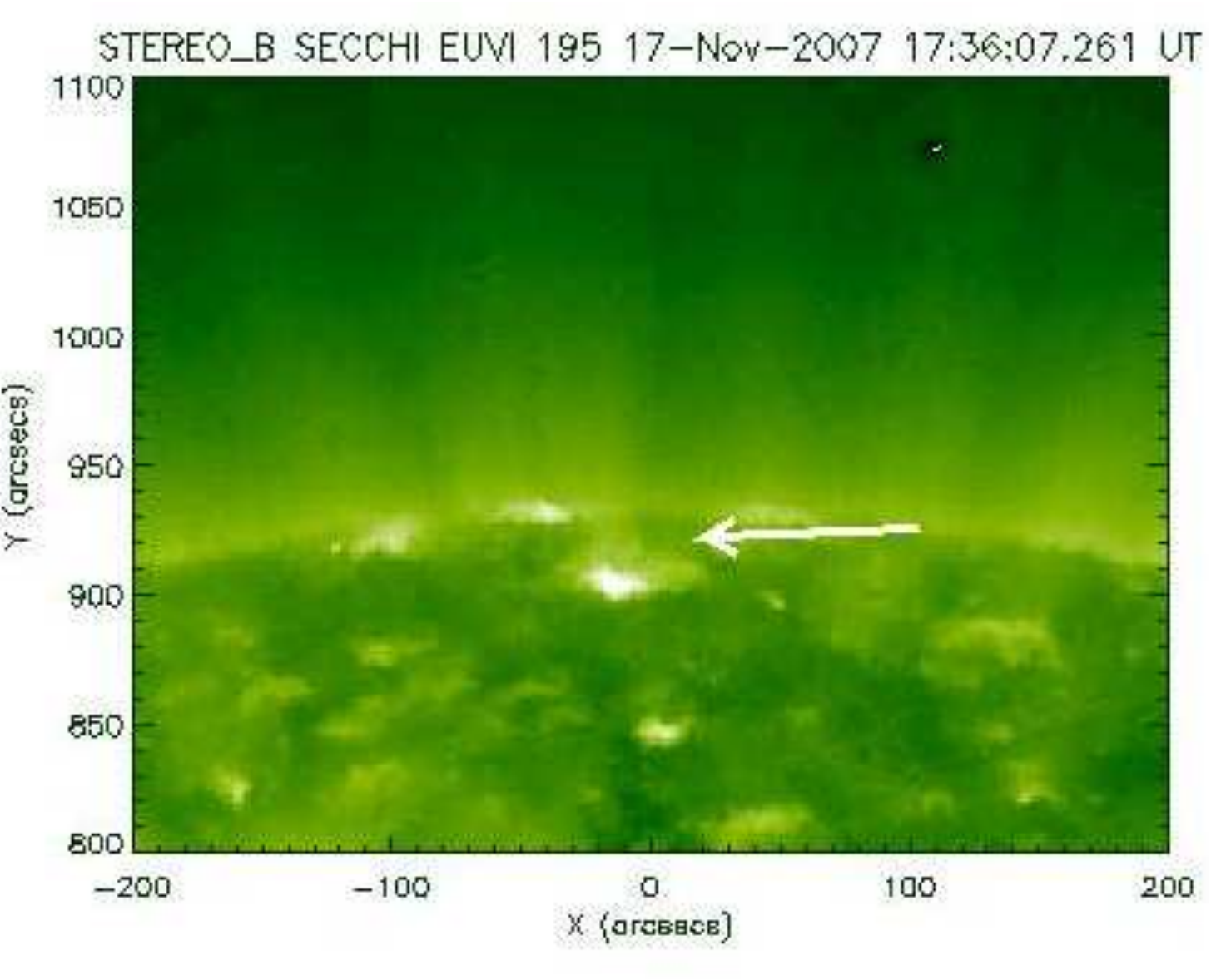} &  
        \includegraphics[width=5.6 cm]{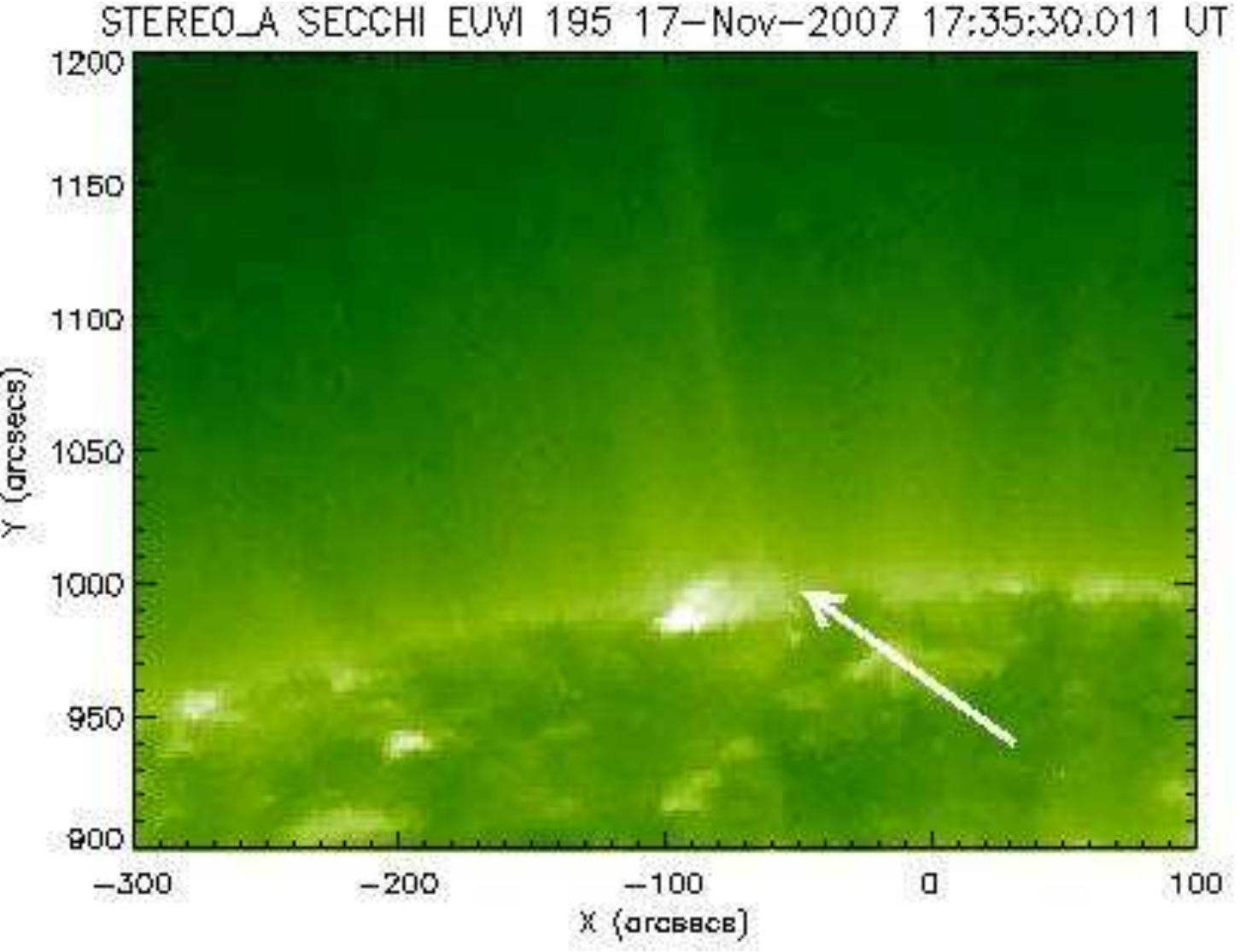}\\
        \end{tabular}
 \caption{ Same ``$\lambda$'' type coronal jet event as in Fig. 4, but seen at 195 \AA~by STEREO B (left) and A (right) on November 17, 2007.}
\label{fig_5}
\end{figure}

Fig. \ref{fig_4} shows a sequence of images at 284 \AA~for a north polar jet taken on November 17, 2007, when both spacecraft were separated by $\Delta \phi_{AB} \simeq 40.57^{\circ}$ (event no. 51). In STEREO EUVI A images, the jet is associated with the presence of a bright point at one leg of the loop and the ejection is developed at the opposite leg. In STEREO EUVI B only the presence of the bright point is visible. For this same event, images in 195 \AA, separated in time of one minute with respect to images at 284 \AA, show ejection at the same heliographic position but the footpoint seems to be characterized by a small loop (Fig. \ref{fig_5}). This event shows that the jet features, and even its very detection, can change with the point of view and with different wavebands.
Moreover, images from STEREO A clearly show a shift from left to right in the jet position. Such a shift suggests an evolution of the magnetic reconnecting region, maybe a ``kink'' instability, which could be compared with the results of numerical simulations ({\it e.g.} \opencite{Yokoyama96}, \opencite{Moreno-Insertis08}, \opencite{Pariat09}) and other data analysis ({\it e.g.}  \opencite{Savcheva07}, \opencite{Filippov09}).
As can be seen from Fig. \ref{fig_3} and Fig. \ref{fig_4}, the jet footpoints typically appear as low coronal bright points or loops. These features sometimes seem to coexist whereas in other cases they appear from different perspectives as the same structure in the SECCHI A and B images. We interpret this finding as projection effects of small loops resembling coronal bright points at lower coronal altitudes when viewed edge on. These observations emphasize the importance of different perspectives.

\begin{figure}[htbp]
     \begin{center}
       \begin{tabular}[htpb]{l l}
        \includegraphics[width=5.6 cm]{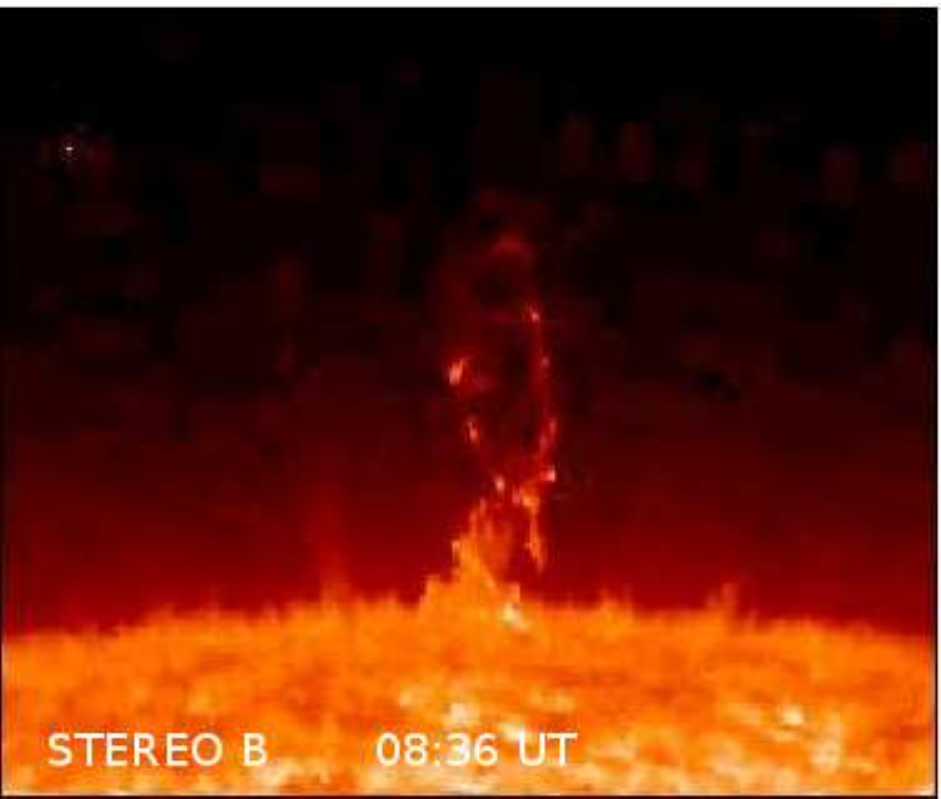} &
        \includegraphics[width=5.6 cm]{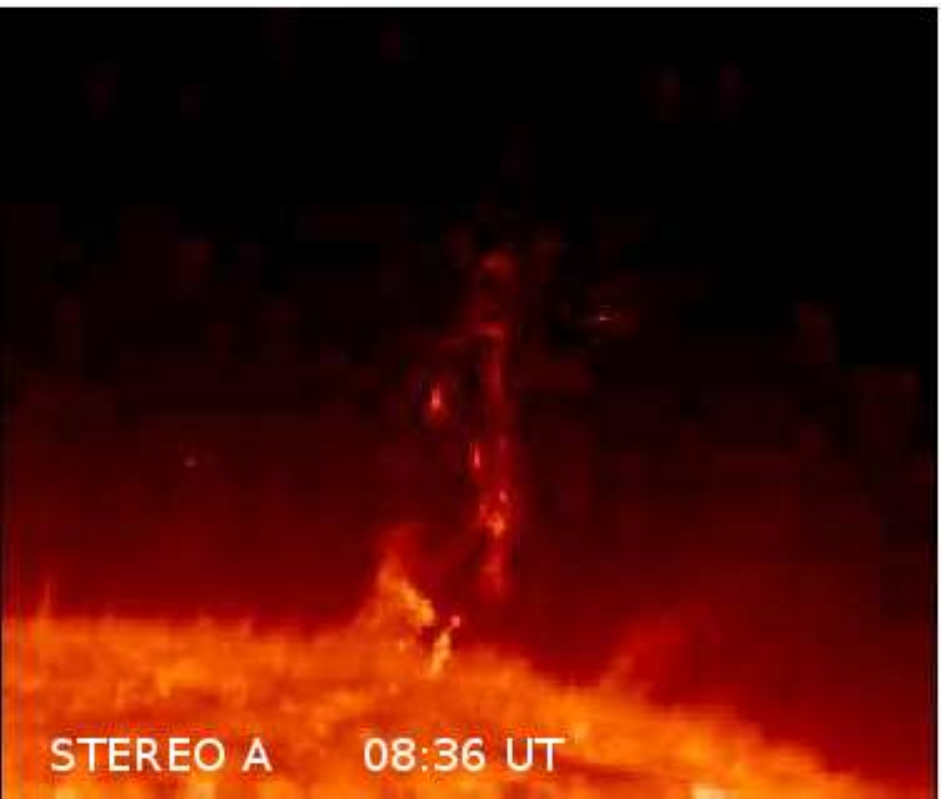}\\

        \end{tabular}
     \end{center}
     \caption{The helical jet event on February 08, 2008. Spacecraft angular separation: $\Delta \phi_{AB} = 45.45^{\circ}$.}
   \label{fig_6}
  \end{figure}

\begin{figure}[htbp]
     \begin{center}
       \begin{tabular}[htpb]{l l}
        \includegraphics[width=5.6 cm]{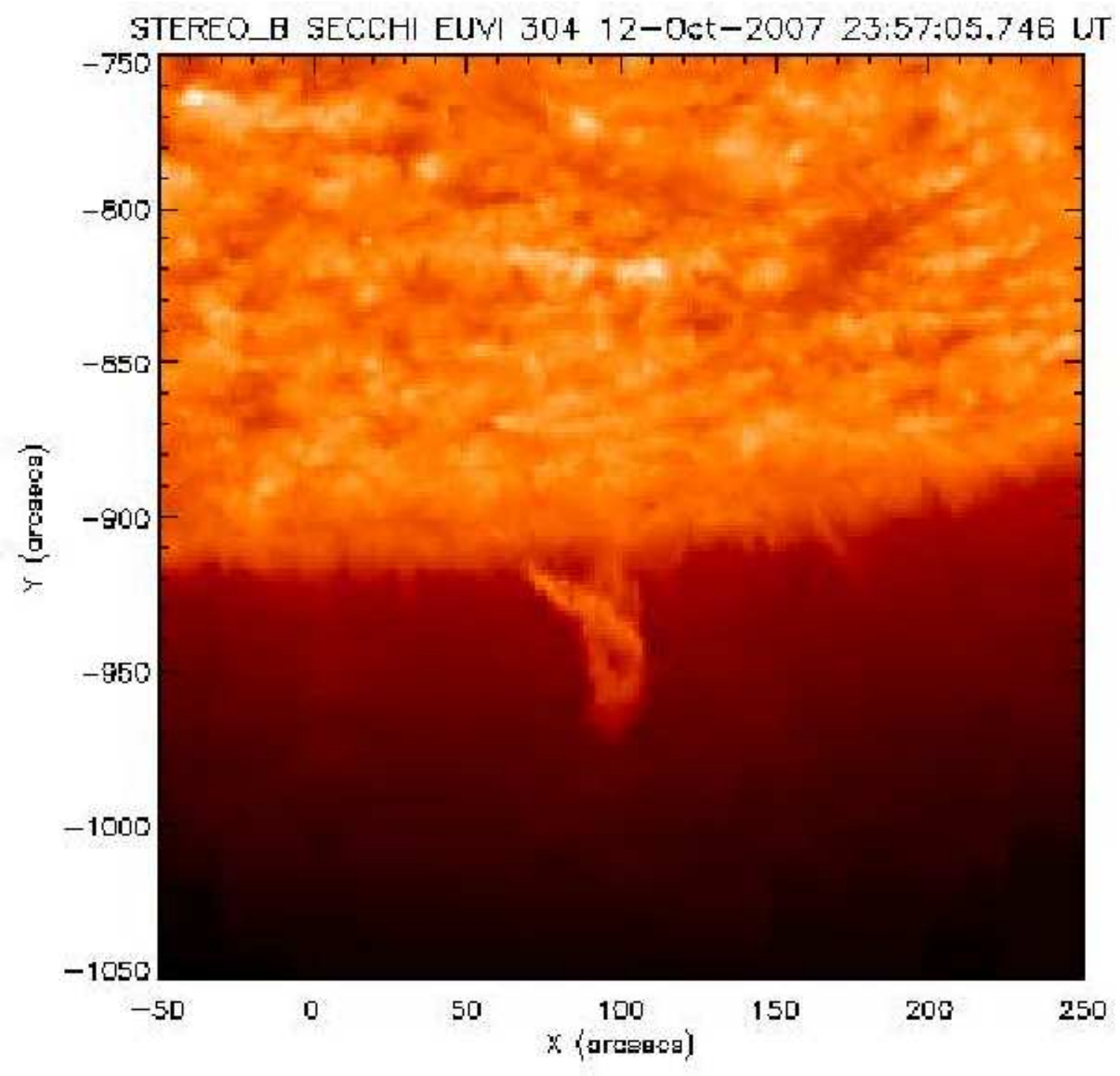} &
        \includegraphics[width=5.6 cm]{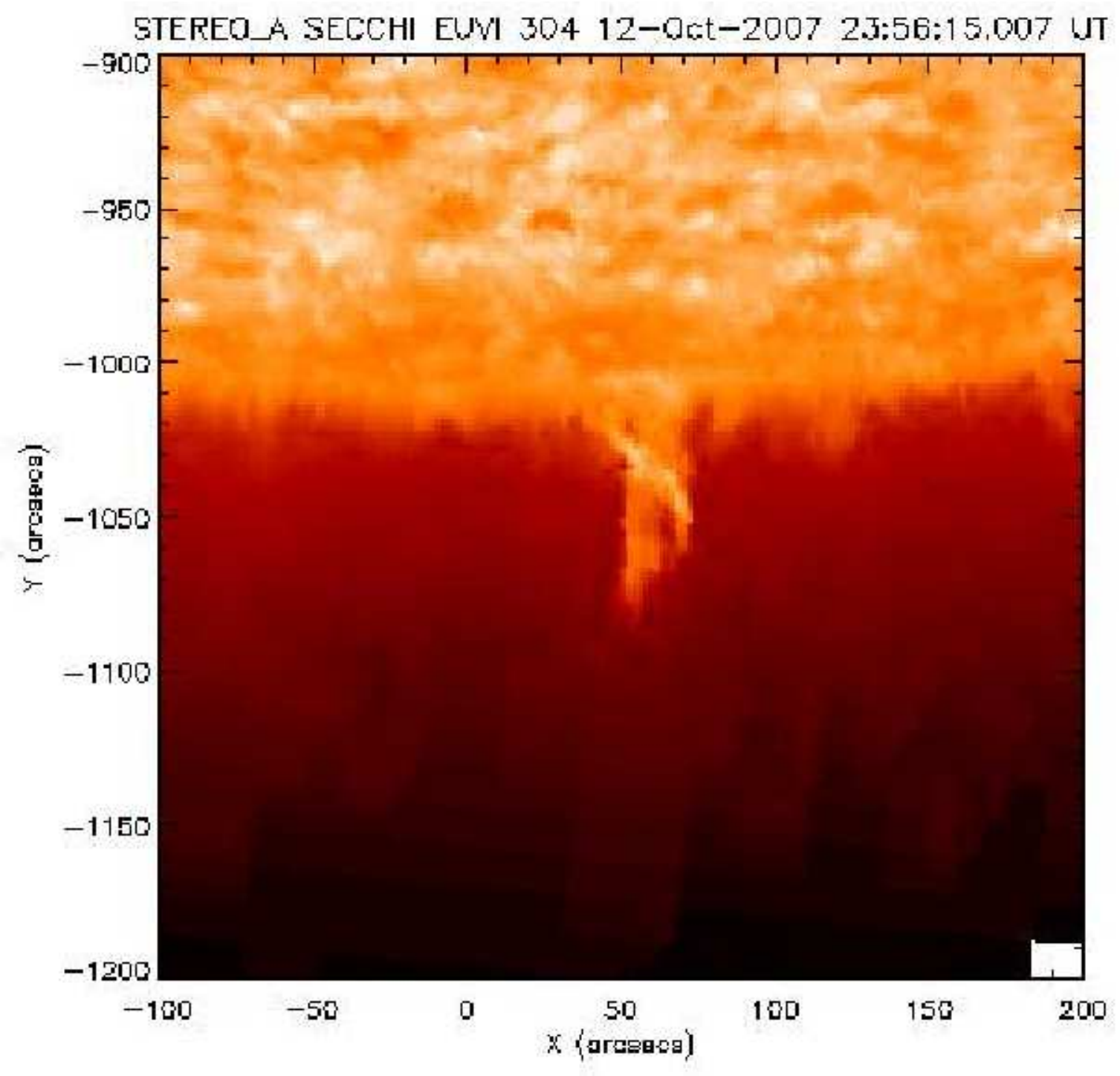}\\

        \end{tabular}
     \end{center}
     \caption{Images of twisted  prominence material in the trailing portion of a micro-CME observed on October 12, 2007. Spacecraft angular separation: $\Delta \phi_{AB} \simeq 35.84^{\circ}$.}
   \label{fig_7}
  \end{figure}

Another important feature of the jets, which we investigated, was whether a helical structure along the jet axis, as first reported by \cite{Patsourakos08}, could be identified. A sample event is presented in Fig. \ref{fig_6}.
Out of the total number of 79 events, 31 events clearly revealed a helical structure, as indicated in the Table of the jet events in the Appendix. Fig. \ref{fig_6} shows a north polar jet which also exhibited a prominent helical structure but not seen to originate from an emerging closed loop system in 171/195 \AA. It was observed at 304 \AA~ on February 8, 2008 (event no. 75), when the angular separation was $\Delta \phi_{AB} = 45.45^{\circ}$. It can be seen that the two helical arms appear well separated in STEREO B, while they partly overlap in STEREO A. This helical structure is also clearly seen at 171 \AA~ and 195 \AA, although the images are less bright. Since the presence of a helical structure is important to test the validity of different jet models ({\it e.g.} \opencite{Pariat09}) these events are important cases for further modeling.
The question why not all jets show helical structure cannot be uniquely answered. A possibility is that they were very narrow so that the twist could not be resolved, as supported by the cases when the jet widths were similar from both perspectives ({\it e.g.} events no. 20--21--29--67--73).
This is what one expects for jets that, to first order, are azimuthally-symmetric structures, {\it i.e.} invariant by rotation, so that they show the same width from different viewpoints. On the contrary, a slab-like geometry would lead to significantly different widths for different perspectives. A detailed study of these characteristics would provide further insight into the validity of 3D models invoking magnetic structures which may  be axially symmetric or not \cite{Moreno-Insertis08, Pariat09}.
A surprise during the study of the jet morphologies was to find events that revealed the same morphology as typically large-scale three part structured CMEs, consisting of a bright leading edge, a dark void and bright trailing edge (being the prominence material) but on much smaller scales \cite{Cremades04}. These so-called micro-CME events are described in more detail in the paper by \inlinecite{Bothmer09}. Another sample jet event (no. 41 of the catalogue) showing the appearance of a twisted small-scale prominence at 304 \AA~arising from inside the south coronal hole on October 12, 2007 is presented in Fig. \ref{fig_7}. Such twisted prominences are well observed features on much larger scale (\url{http://sohowww.nascom.nasa.gov/gallery/bestofsoho.html}). The spacecreaft's angular separation at that time was $\Delta \phi_{AB} \simeq 35.84^{\circ}$. The difference in the viewpoints is obvious and confirms that it is a real twisted 3D loop.
It seems likely that the overall magnetic topology depends on the source region characteristics of the underlying photospheric magnetic field as in the case of large-scale CMEs found to arise from bipolar magnetic field regions on larger scale as shown by \inlinecite{Cremades04}. Similar to the large-scale CMEs, the micro-CMEs may also evolve from bipoles with enhanced magnetic flux compared to the surrounding fields but on much smaller scales. This is similar to the observation that coronal jets seem to arise close to small bipoles within the coronal holes as shown by \inlinecite{Shimojo96}. However it is beyond the scope of this study to fully investigate the coupling of the coronal and photospheric structures. 
Overall we found evidence for 5 micro-CMEs in the total set of 79 events.

The list of coronal jet events presented in the Appendix provides information on the identified morphology of the individual events. 
However, it should be pointed out that for a number of events a unique classification was not possible, because sometimes the jet features were different in each wavelength and sequences at different time cadences had to be investigated; in the Appendix this is indicated by a question mark.

\section{Coronal jet lifetimes}

\begin{figure}[htbp]
     \begin{center}
       \begin{tabular}[htpb]{l l}
        \includegraphics[width=5.6 cm]{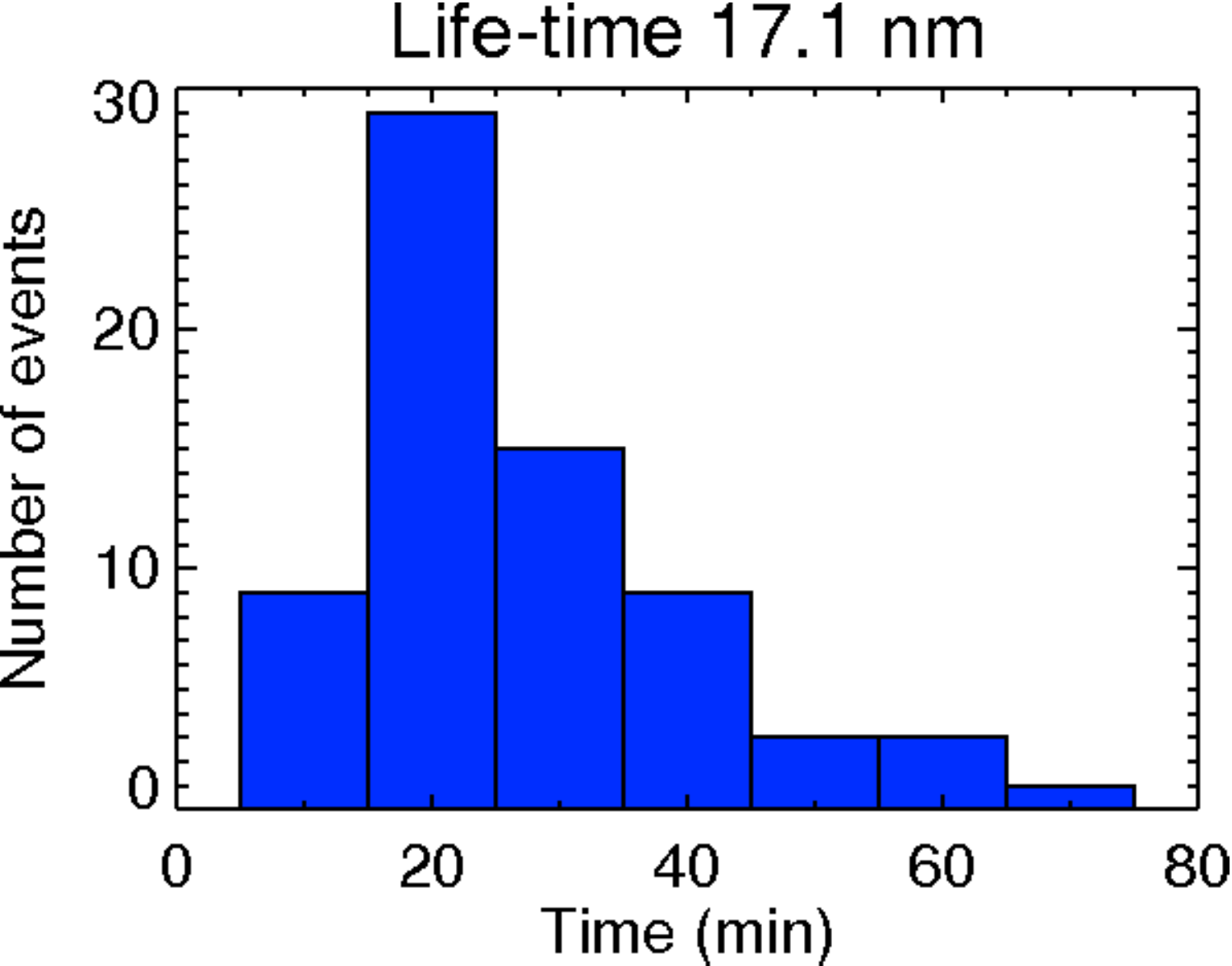} &
        \includegraphics[width=5.6 cm]{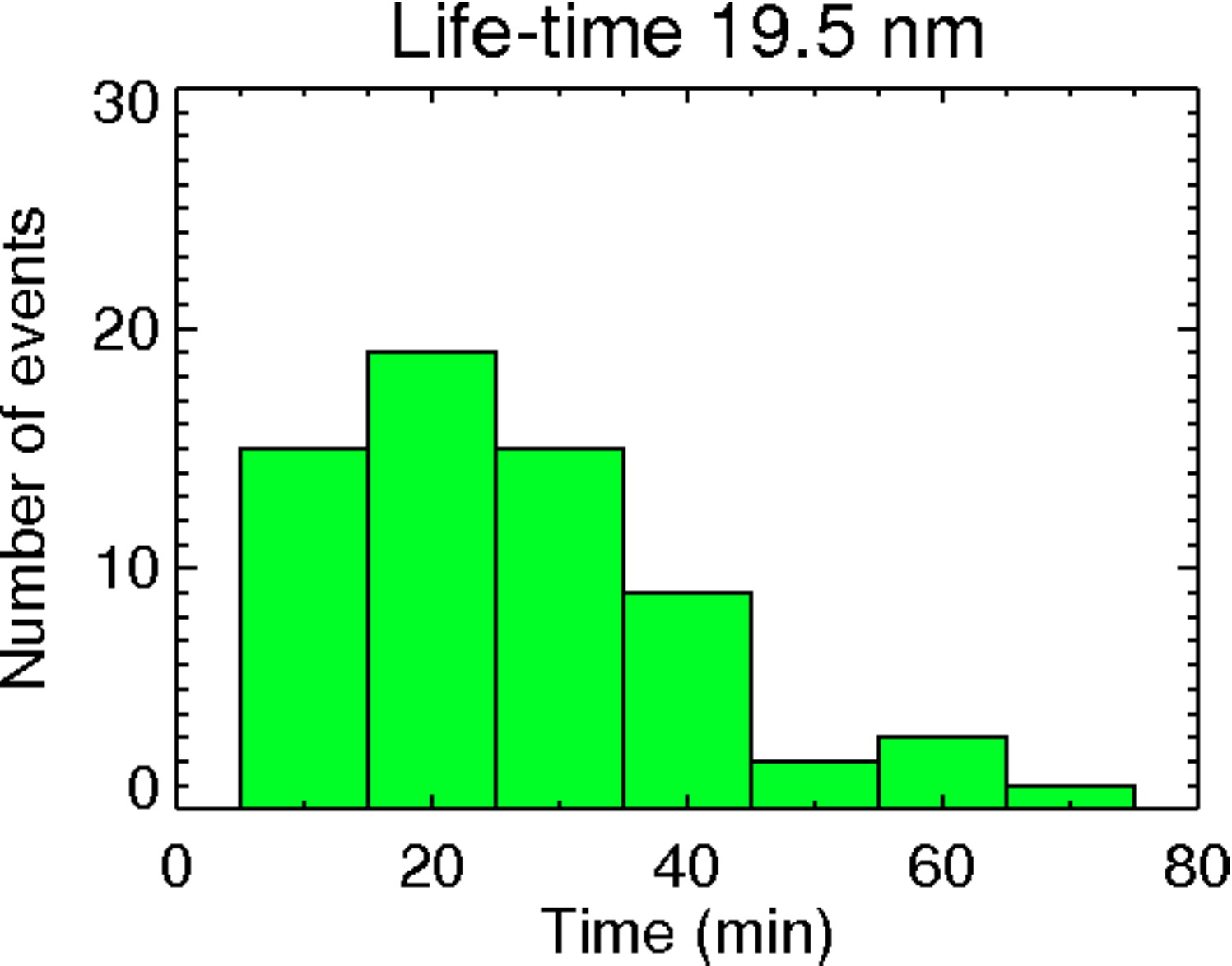}\\
        \includegraphics[width=5.6 cm]{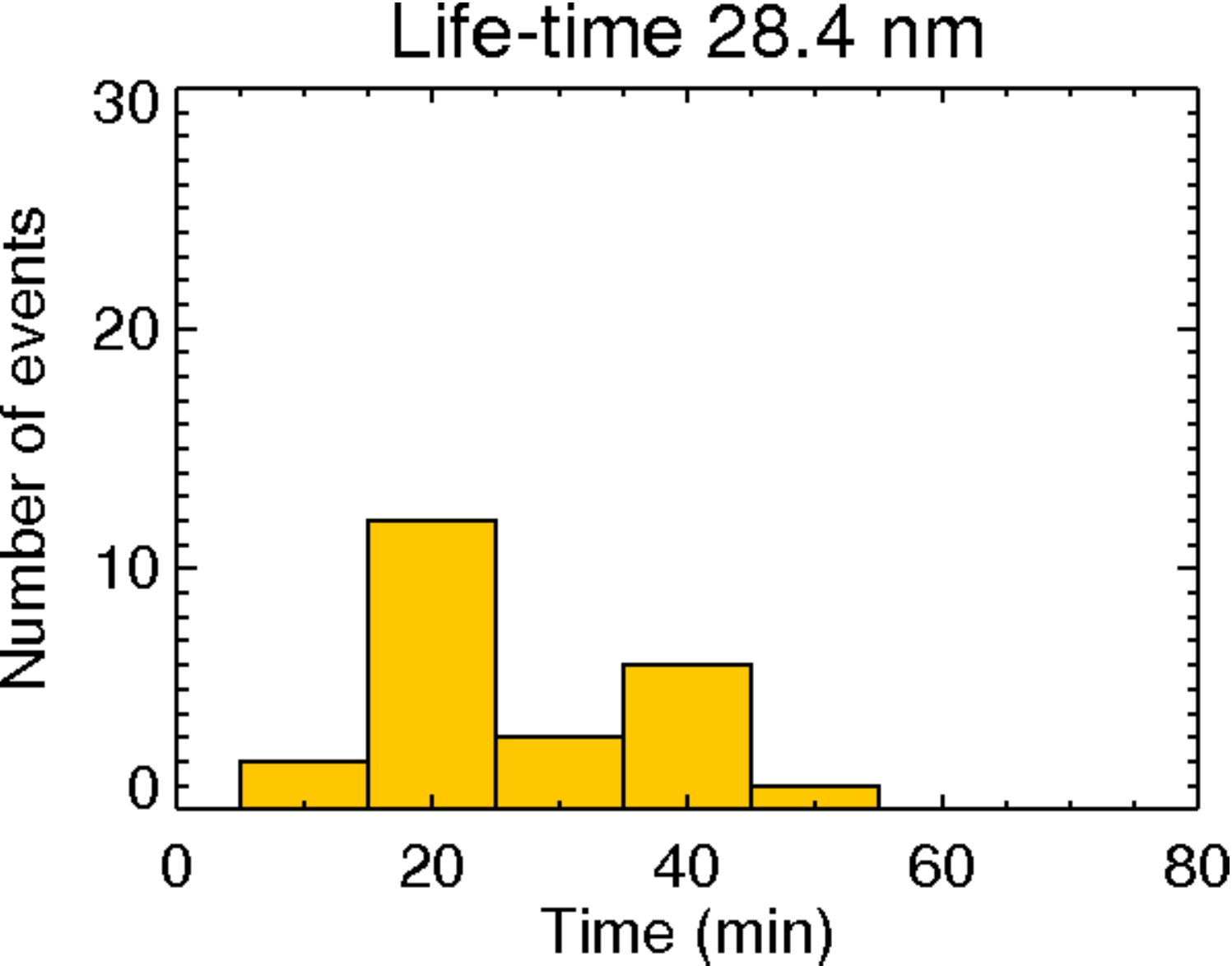} &
        \includegraphics[width=5.6 cm]{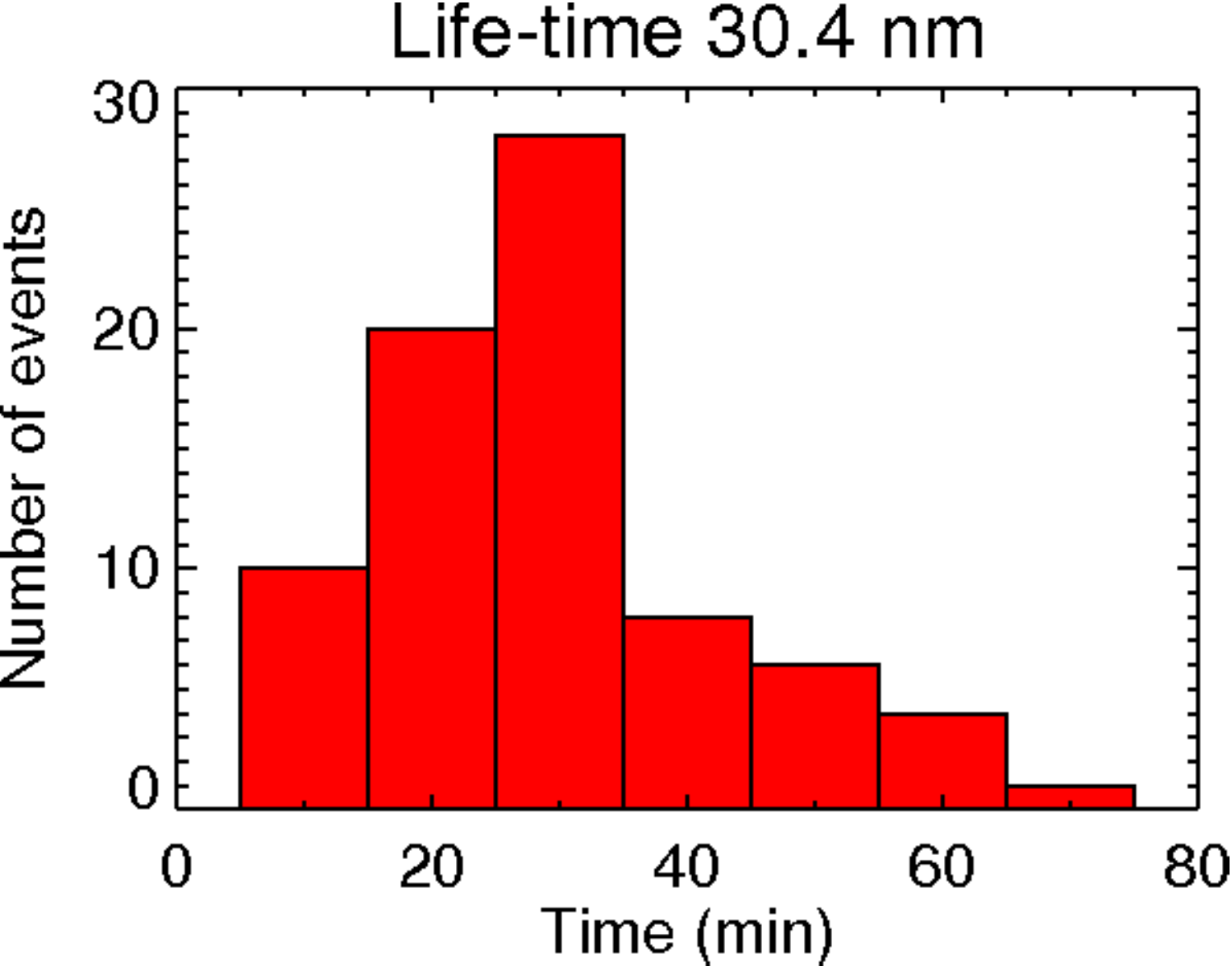}\\
        
        \end{tabular}
     \end{center}
     \caption{Distribution of the lifetime of coronal jets seen by STEREO/SECCHI at 171, 195, 284 and 304 \AA. Note the shift of the peak in the lifetime distribution at 304 \AA.}
   \label{fig_8}
  \end{figure}

Inspection of the polar jets reported in the catalogue also allows to obtain some basic information about the duration of observation in the EUVI field of view, {\it i.e.}, about the lifetime of the jets at the different wavelengths in the low corona. These EUVI visibilities correspond to the lifetimes given in the third column of the catalogue for each wavelength. A statistical distribution for the lifetimes of all jets at the different wavelengths between 1--1.7 R$_\odot$ in bins of 10 minutes is presented in Fig. \ref{fig_8}. 

\begin{figure}[htbp]
     \begin{center}
       \begin{tabular}[htpb]{l l}
        \includegraphics[width=5.6 cm]{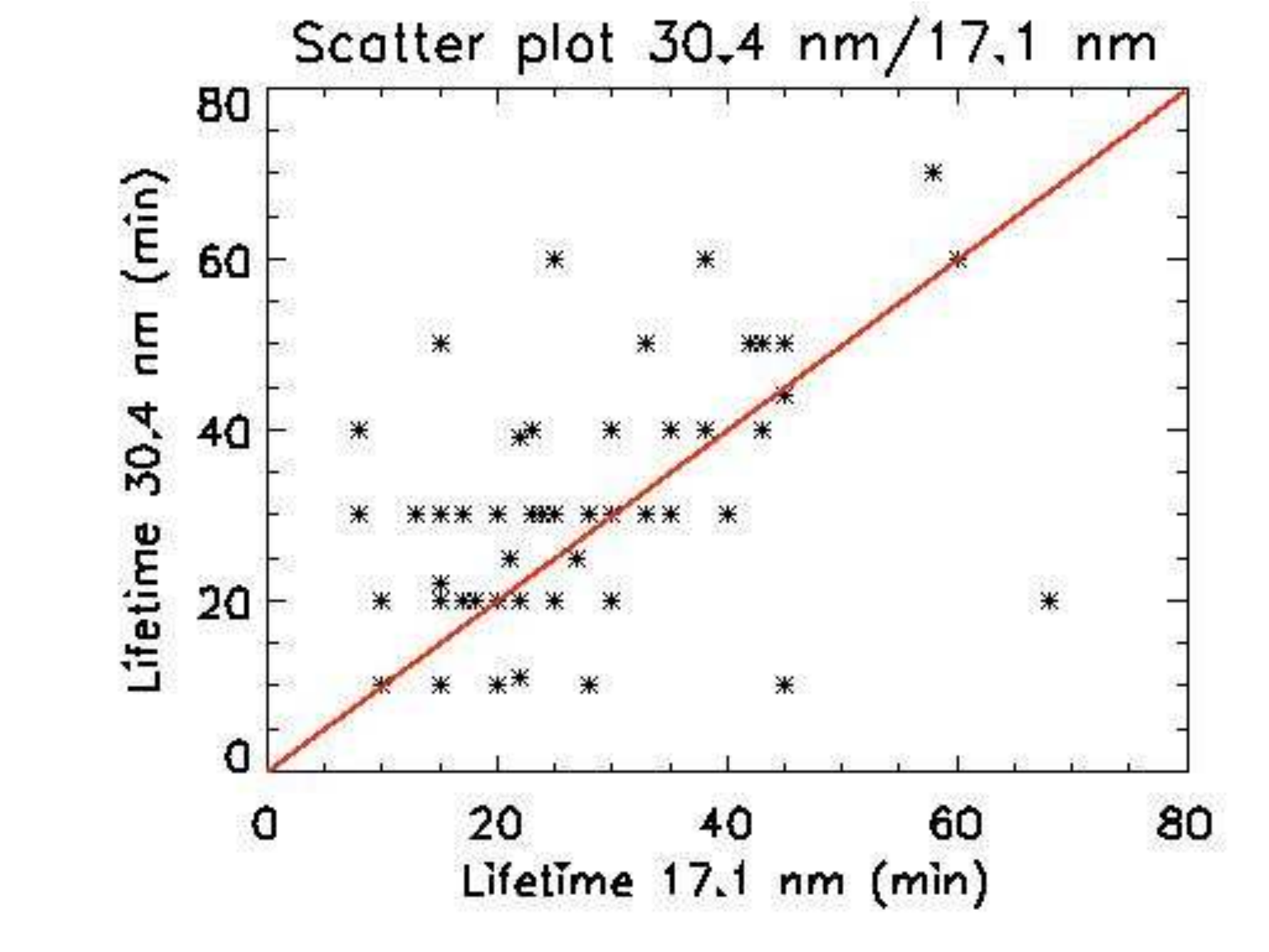} &
        \includegraphics[width=5.6 cm]{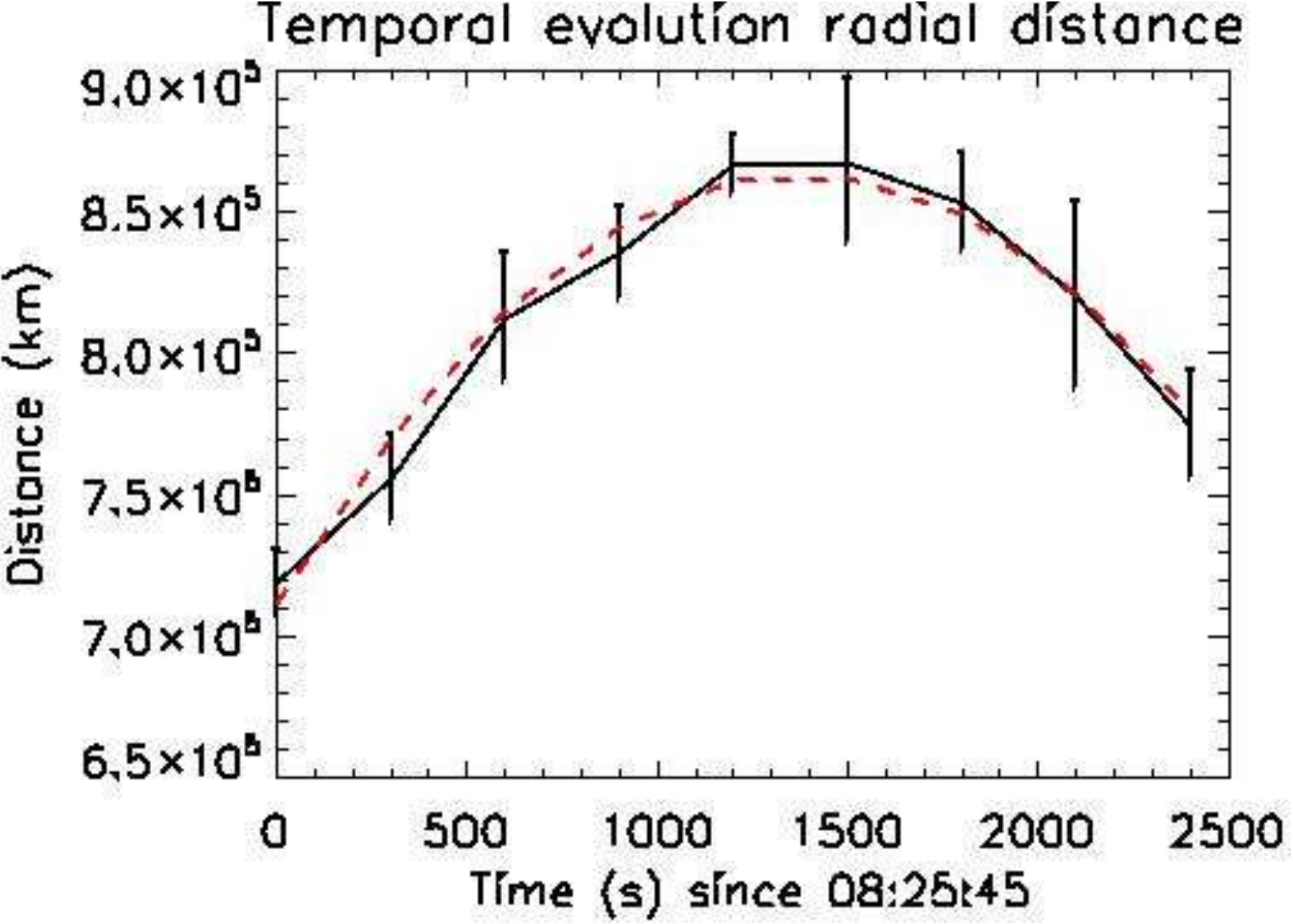}  \\ 
       \end{tabular}
     \end{center}
     \caption{Left: scatter plot of 304 \AA~and 171 \AA~lifetimes. The majority of events shows longer duration in 304 \AA~than in 171 \AA. Right: a plot of the leading edge position vs time for the micro-CME of 6 May, 2007, at 304 \AA. We can obtain the starting speed of leading edge jet ($v_0=(219.86\pm22.38)~km/s$) and the acceleration ($a=(-0.1596\pm0.0198)~km/s^2$) by a quadratic fit (red dashed line).}
   \label{fig_9}
  \end{figure}

It should be noted that the estimated lifetime is influenced by the different time cadence with which the EUVI telescope operates: usually a time cadence of 2.5 min is used for 171 \AA, while the cadence is 10 min for 195 \AA$~$ and 304 \AA. It can be seen from Fig. \ref{fig_8} that the lifetimes of the jets analysed in this study at 171 \AA$~$and 195 \AA$~$ are peaked at 20 min; conversely, at 304 \AA$~$, the lifetime distribution is peaked at 30 min.
The lifetime distribution at 284 \AA$~$is peaked at 20 min and 40 min  but the number of events is lower than in the other wavebands; in this case the two peaks may just correspond to multiples of the most used cadence.
The differences in lifetimes emphasizes the fact that looking at different wavelengths implies observing the solar corona at different temperatures: the spectral line at 304 \AA~ corresponds to lower temperatures ($\sim80,000$ K), which  suggests that the duration of the event is usually longer at lower temperatures. On the other hand, the shorter lifetime at 171/195 \AA~ could also be related to the fact that at these wavelengths jets have coronal temperatures ($\sim10^6$ K) and dim very fast with distance, since they travel in the hot corona and contrast is low. Conversely, 304 \AA~jets have larger constrast since there is not so strong emission at this wavelength in the corona. This implies that the lifetimes at 195 \AA~could be longer by at least one exposure. From Fig. \ref{fig_9}a it can be seen that for each event the lifetime is usually shorter at 171 than 304 \AA; this can be compared with the lifetime in the X-rays, which is even shorter, as reported by \inlinecite{Savcheva07} and \inlinecite{Cirtain07}. A very rough estimate of jet speeds can be obtained from the jet lifetime by assuming that the jet speed does not change too much during the transit time in the EUVI field of view of 0.7 R$_\odot$ beyond the limb. This means that 171 \AA~plasma has an estimated outward propagating speed of about 400 km/s, whereas 304 \AA~cooler material is moving at lower speeds of about 270 km/s. If we accept these speed estimates as typical of the corresponding jets, we can infer that 304 \AA~ cooler  material is not capable to reach a speed sufficient to leave the Sun at the given height, so that material is falling back (similar to spicula matter), as can be noted in some events ({\it e.g.} no. 14--15--71--73) (the escape speed from the Sun at the distance of 1.7 R$_\odot$ is $\sim$360 km/s).

A more definitive way to measure the jet speed is by determining the leading edge jet position in 3D for a number of consecutive frames. This has been done with the help of the \emph{SolarSoft} routine (\texttt{scc\_triangulate}) and is shown for the event no. 15 of the catalogue in Fig. \ref{fig_9}b. A parabolic fit yields a maximum jet speed of $\sim$200 km/s, consistent with the above estimates at 304 \AA. Also, the inferred (sunward) acceleration is less than the solar gravity, which shows that electromagnetic forces are still acting on the jet material. Other 3D velocity determination will be presented elsewhere.
 
An estimate of the lifetimes can also be obtained from the coronagraph observations. In COR1 A the peak of the distribution is centered near 70 min, while in COR1 B the peak of the distribution is centered near 80 min. A lifetime of 75 minutes for the field of view of COR1 corresponds to a speed of about 390 km/s which is in good agreement with the speed derived for the hot 171 \AA~ plasma components of the jets. The fact that the estimated lifetime in COR1 B is higher by ten minutes might be explained by its better stray-light rejection since this feature allows to trace jets farther away.
 \section{Conclusions}

In this work we presented the first comprehensive statistical study of polar coronal jets with the SECCHI instrument package on the STEREO spacecraft, using EUVI A and B and COR1 A and B data. A catalogue of 79 jets identified from simultaneous EUVI and COR1 coronagraph observations by both spacecraft has been compiled. The observations were taken during a period at solar activity minimum from March 2007 to April 2008.
From a systematic investigation of the 79 events observed by both spacecraft at separation angles between 2 and 48 degrees we find the following basic properties of coronal jets:

\noindent$\bullet$ 37 \emph{Eiffel tower} type jets in which jets show  a shape similar to an inverted-Y and they are associated to the magnetic configuration of a small magnetic bipole reconnecting with the ambient coronal field at its looptop;

\noindent$\bullet$ 12 \emph{lambda} type jets in which ejection is observed to be shifted from the position of a bright point or a small loop, and this topology is associated to the magnetic configuration of a small bipole reconnecting with the unipolar coronal field at its footpoint; 

\noindent$\bullet$ besides the previous two morphologies, there are 5 micro-CME type events \cite{Bothmer09} characterized by the evidence of a small loop that elongates from the solar surface and resemble the CMEs but on smaller scale.
 
Without more sophisticated analyses of jets it is difficult to provide a reliable interpretation why we have identified about three times more ET than $\lambda$ events. ET events at looptops are likely have better visibility against the background compared to reconnection processes happening at the footpoints because those appear bright and heating processes are less visible. Another plausible interpretation would be that ET-type jets could require some build-up of magnetic energy and then an instability like discussed by \inlinecite{Patsourakos08}, \inlinecite{Pariat09}. Therefore, whenever they occur more energy has been stored and therefore released and the jets are more bright, thus more detectable. On the other hand, $\lambda$ jets can occur more ``easily'' without so much energy having been stored, and therefore are less observable, even though they may be more numerous. 
 
Overall there are 31 events that clearly exhibit a helical structure of the magnetic field.
It is important to note that these findings imply that a jet observed by a coronagraph at heights of $\sim$1.5 $R_\odot$ can have different low coronal onset scenarios in terms of the magnetic fields structure and its evolution.

The typical lifetimes in the EUVI FOV are 20 minutes at 171 \AA~, 30 minutes at 304 \AA, while in COR1 the lifetimes are peaked at around 70--80 minutes. The corresponding estimated speeds with respect to the FOV of the EUVI and COR1 telescopes are 400 km/s for the hot 171 \AA~plasma component and only 270 km/s for the cooler 304 \AA~chromospheric component. The speed of 400 km/s is comparable to that derived from the COR1 FOV of 390 km/s. 

The present catalogue can serve for several purposes. One is to undertake a more detailed study of specific events to determine the 3D velocity and acceleration and to carry out a 3D reconstruction of the jet, to compare with numerical simulations. A second purpose is to look for the association between jets and other chromospheric phenomena, like spiculae, plumes, etc., \cite{Raouafi08} which is useful for constraining more complete physical models of the solar activity in the coronal holes. A third purpose is to fill the gap between large scale coronal phenomena like CME and the small scale phenomena in polar coronal holes.   

\clearpage
\appendix

In this appendix, polar jet events are organized in a catalogue. Each event is labeled by  progressive number, date of observation, angular separation $\Delta \phi_{AB}$ between STEREO A and STEREO B, time of visibility in EUVI and in COR1, position angle in EUVI ($\alpha$) and in COR1 ($\beta$), brief indication on the morphology of the event and presence of helicity. The position angles are given for STEREO A, except when differently noted. 
The following abbreviations are used:
\textbf{ET}: Eiffel tower;
\textbf{Lambda}: lambda jet;
\textbf{N(S)PCH}: North (South) polar coronal hole;
\textbf{L}: limb;
\textbf{I}: inside;
\textbf{EDGE}: edge of coronal hole;
\textbf{Hel}: evidence of helical structure.
The question mark \textbf{?} indicates unclassified jets events.

\section{Catalogue of polar coronal hole jets.}
\tiny
\begin{longtable}{ p{0.4 cm}  p{1.3cm} l p{3 cm} p{2.4 cm}}

\hline
\bfseries{N$^\circ$} &\bfseries{Date}  &\bfseries{EUVI}      &\bfseries{COR1}    &\bfseries{Position}  \\ 
                     &\bfseries{$\Delta \phi_{AB}$}&                &          & \bfseries{Morphology}\\
\hline                      
 1   &  2007-03-19 &  171   16:08-16:38 &     A:16:20-17:30  & SPCH, L \\           
     &             &  195   16:01-16:31 &     B:16:20-17:20  & $\alpha(A)=164^\circ$ \\
     &$2.16^\circ$ &  284   16:08-16:38 &                    & $\beta(A)=158^\circ$\\
     &             &  304   16:01-16:31 &                    & \bfseries{?}              \\
\hline 
 2   &  2007-03-19 &  171   16:08-16:28 &     A:16:10-17:20  & SPCH, L\\  
     &             &  195   16:01-16:31 &     B:16:20-17:30  & $\alpha(A)=176^\circ$\\
     &$2.16^\circ$ &  284   15:58-16:28 &                    & $\beta(A)=175^\circ$\\              
     &             &  304               &                    & \bfseries{?}              \\
\hline
 3   &  2007-03-20 &  171               &     A:            & NPCH, EDGE\\
     &             &  195               &     B: 08:51-09:31&  $\alpha(B)=6^\circ$\\
     &$2.20^\circ$ &  284               &                   &  $\beta(B)=6^\circ$ \\
     &             &  304   08:31-09:01 &                   &  \bfseries{Lambda}              \\
\hline
 4   &  2007-03-24 &  171   00:38-00:58 &     A:01:10-02:20 & SPCH, L\\
     &             &  195   00:31-00:51 &     B:01:10-02:30 &  $\alpha(A)=181^\circ$\\
     &$2.44^\circ$ &  284   00:48-00:58 &                   &  $\beta(A)=182^\circ$ \\
     &             &  304   00:51 (very faint)&             &  \bfseries{ET}              \\ 
\hline    
 5   &  2007-03-27 &  171   not very clear&   A:22:10-23:50 & SPCH, L  \\
     &             &  195   21:41-22:01 ? &   B:22:10-23:30?& $\alpha(A)=176^\circ$\\
     &$2.72^\circ$ &  284   21:58-22:08   &                 & $\beta(A)=167^\circ$  \\
     &             &  304   21:51-22:11   &                 & \bfseries{?}            \\
\hline
 6   & 2007-03-30  &  171   14:49-15:01    &  A:15:20-16:30 & NPCH, I  \\
     &             &  195   14:42-14:52    &  B:15:10-16:30 & $\alpha(A)=1^\circ$\\
     &$2.92^\circ$ &  284   14:40-15:00    &                & $\beta(A)=2^\circ$\\
     &             &  304   14:51          &                & \bfseries{ET}               \\
\hline
 7   & 2007-04-01  &  171   14:24-14:54    & A:14:40-16:00? & SPCH, L  \\
     &             &  195   14:22-14:52    & B:14:50-15:30? & $\alpha(A)=170^\circ$  \\
     &$3.08^\circ$ &  284                  &                & $\beta(A)=166^\circ$\\
     &             &  304   14:31-14:51    &                & \bfseries{Lambda}               \\
\hline
 8   & 2007-04-08  & 171   15:09-15:31     & A:15:30-16:30  & SPCH, I  \\
     &             & 195   15:12-15:22     & B:15:40-16:50  & $\alpha(A)=192^\circ$ \\
     &$3.68^\circ$ & 284                   &                & $\beta(A)=200^\circ$\\
     &             & 304   15:11-15:22     &                & \bfseries{Lambda}               \\
\hline
 9   & 2007-05-01  & 171   11:19-11:34     & A:11:20-12:20  & SPCH, L  \\
     &             & 195   11:22-11:42     & B:11:30-12:15  & $\alpha(A)=183^\circ$ \\
     &$6.15^\circ$ & 284                   &                & $\beta(A)=186^\circ$\\
     &             & 304   11:11-11:41     &                & \bfseries{?}               \\
\hline
10   & 2007-05-01  & 171   18:04-18:29     & A:18:30-20:00  & NPCH, L  \\
     &             & 195   18:12-18:32     & B:18:20-20:00? & $\alpha(A)=0^\circ$ \\
     &$6.18^\circ$ & 284   18:20           &                & $\beta(A)=1^\circ$\\
     &             & 304   18:11-18:31     &                & \bfseries{ET-Hel}               \\
\hline
11   & 2007-05-02  & 171   13:01-13:16     & A:13:10-14:30  & SPCH, EDGE\\
     &             & 195   13:02-13:22     & B:13:15-  ?    & $\alpha(A)=153^\circ$ \\ 
     &$6.28^\circ$ & 284   13:00           &                & $\beta(A)=152^\circ$\\ 
     &             & 304   13:01-13:51     &                & \bfseries{ET-Hel}               \\
\hline
12   & 2007-05-02  & 171   20:59-21:21     & A:21:20-22:10  & NPCH, L  \\
     &             & 195   21:02-21:22     & B:21:10-22:35  & $\alpha(A)=0^\circ$ \\  
     &$6.33^\circ$ & 284                   &                & $\beta(A)=1^\circ$\\ 
     &             & 304   21:01-21:31     &                & \bfseries{ET-Hel}               \\
\hline
13   & 2007-05-04  & 171    09:48-10:01    & A:             & SPCH, L     \\   
     &             & 195    09:46-10:01    & B:             & $\alpha(A)=177^\circ$ \\
     &$6.52^\circ$ & 284    09:46-10:01    &(diff.movie(A):10:15)&$\beta(A)=175^\circ$   \\
     &             & 304    09:50-10:20    &                 &\bfseries{Lambda}   \\
\hline
14   & 2007-05-05  & 171    04:03-04:18    & A:             & NPCH, L    \\         
     &             & 195    04:06-04:21    & B:             & $\alpha(B)=16^\circ$ \\  
     &$6.62^\circ$ & 284    04:06          & (diff.movie(B):04:25-&$\beta(B)=26^\circ$ \\
     &             & 304    04:05-04:35    & 05:05)         & \bfseries{Micro-CME}\\
\hline 
15   & 2007-05-06  & 171   08:26-08:48     & A:08:40-09:15  & SPCH, L \\
     &             & 195   08:26-08:56     & B:08:55-09:40  & $\alpha(A)=172^\circ$  \\
     &$6.78^\circ$ & 284   08:26-08:56     &                & $\beta(A)=167^\circ$\\ 
     &             & 304   08:31-09:10     &                & \bfseries{Micro-CME}              \\
\hline
16   & 2007-05-07  & 171                   & A:15:30-16:30  & SPCH, L               \\   
     &             & 195   14:46           & B:difficult    & $\alpha(A)=176^\circ$ \\
     &$6.94^\circ$ & 284                   & better seen    & $\beta(A)=177^\circ$\\
     &             & 304   14:50-15:05     & in diff.movie & \bfseries{?}              \\
\hline
17   & 2007-05-09  & 171   10:40           & A:10:50-11:50? & NPCH, W,    L \\
     &             & 195   10:41           & B:10:40-11:50? & $\alpha(A)=346^\circ$ \\
     &$7.19^\circ$ & 284   10:41           &                & $\beta(A)=316^\circ$\\
     &             & 304   10:35-11:06     &                & \bfseries{?}              \\
\hline
18   & 2007-05-10  & 171                   & A:08:10-09:30  & SPCH, I \\  
     &             & 195                   & B:             & $\alpha(A)=175^\circ$ \\
     &$7.31^\circ$ & 284                   &                & $\beta(A)=187^\circ$\\
     &             & 304   07:33-07:51     &                & \bfseries{?}              \\
\hline        
19   & 2007-05-11  & 171    17:04-17:31    & A:             & NPCH, L/EDGE\\
     &             & 195    17:06-17:26    & B:             & $\alpha(A)=346^\circ$ \\
     &$7.50^\circ$ & 284    17:11-17:26    & (diff.movie(A):17:45)& $\beta(A)=333^\circ$\\
     &             & 304    17:10-17:35    &                & \bfseries{?}              \\
\hline
20   & 2007-05-24  & 171    09:28-09:46    & A:09:55-10:35  & NPCH, L  \\ 
     &             & 195    09:25-09:55    & B:09:55-10:35  & $\alpha(A)=7^\circ$ \\
     &$9.38^\circ$ & 284                   & very difficult observing &$\beta(A)=6^\circ$\\ 
     &             & 304    09:36-09:56    & signal in COR1 & \bfseries{ET}              \\
\hline
21   & 2007-05-24  & 171   21:48-22:16     & A:22:05-23:05  & SPCH, L \\ 
     &             & 195   21:55           & B:22:15-23:05  & $\alpha(A)=169^\circ$  \\
     &$9.46^\circ$ & 284                   &                & $\beta(A)=165^\circ$\\   
     &             & 304   21:46-21:56     &                & \bfseries{ET-Hel?}              \\
\hline
22   & 2007-05-25  & 171   03:18-03:41     & A:02:45-04:45  & SPCH, L \\ 
     &             & 195   03:25           & B:03:05-03:55  & $\alpha(A)=182^\circ$ \\
     &$9.50^\circ$ & 284   03:26           &                & $\beta(A)=185^\circ$\\
     &             & 304   03:16-03:46     &                & \bfseries{ET-Hel}              \\
\hline
23   & 2007-05-27  & 171   19:38-19:58     & A:20:05-21:15? & NPCH, L \\ 
     &             & 195   19:45-19:55     & B:20:05-21:25  & $\alpha(A)=2^\circ$  \\
     &$9.92^\circ$ & 284   19:46           &                &  $\beta(A)=0^\circ$\\
     &             & 304   19:46-19:56     &                & \bfseries{ET-Hel}              \\
\hline
24   & 2007-05-28  & 171   23:46-00:08     & A:00:15-00:45  & SPCH, L \\ 
     &             & 195   23:45-00:15     & B:             & $\alpha(A)=176^\circ$   \\  
     &$10.11^\circ$& 284                   &                & $\beta(A)=175^\circ$  \\ 
     &             & 304   23:56-00:16     &                & \bfseries{?}              \\
\hline
25   & 2007-06-07  & 171   04:58-05:43     & A:05:25-07:05  & NPCH, I  \\
     &             & 195   04:55-05:55     & B:05:25-07:05  &  $\alpha(A)=7^\circ$ \\
     &$11.65^\circ$& 284   05:06-05:46     &                &  $\beta(A)=12^\circ$\\
     &             & 304   05:06-05:56     &                & \bfseries{ET-Hel}       \\
\hline
26   & 2007-06-13  & 171   not clear       & A:             & NPCH, I \\
     &             & 195   19:25-20:15     & B:19:55-20:25  & $\alpha(B)=0^\circ$ \\
     &$12.81^\circ$& 284   19:26           &                &  $\beta(B)=2^\circ$\\  
     &             & 304   19:26-19:46     &                & \bfseries{ET}              \\
\hline
27   & 2007-07-24  & 171   00:58-01:06     & A:01:15-02:05  & NPCH, I \\
     &             & 195   01:05-01:15     & B:01:15-02:35  & $\alpha(B)=359^\circ$ \\ 
     &$20.57^\circ$& 284                   &   very faint   & $\beta(B)=356^\circ$ \\
     &             & 304   00:56-01:26     &   in COR1A       & \bfseries{?-Hel?}              \\
\hline 
28   & 2007-08-04  & 171   20:38-20:48     & A:20:55-22:15  & NPCH, L \\ 
     &             & 195   20:45           & B:               & $\alpha(A)=1^\circ$  \\
     &$22.97^\circ$& 284   20:46           &                & $\beta(A)=0^\circ$\\
     &             & 304   20:36-20:46     &                & \bfseries{ET}              \\
\hline
29   & 2007-08-13  & 171   01:08-01:26     & A:01:35-02:45  & NPCH, I \\
     &             & 195   01:15-01:25     & B:01:35-03:06  & $\alpha(A)=356^\circ$ \\
     &$24.63^\circ$& 284   01:06           &                & $\beta(A)=351^\circ$\\
     &             & 304   01:16-01:36     &                & \bfseries{ET}              \\
\hline
30   & 2007-09-03  & 171   08:36-09:01     & A:             & NPCH, L \\
     &             & 195   08:35-08:55     & B:08:55-10:05? & $\alpha(B)=1^\circ$ \\
     &$28.85^\circ$& 284   08:46           &                & $\beta(B)=359^\circ$\\
     &             & 304   08:36-09:06     &                & \bfseries{ET-Hel}              \\
\hline
31   & 2007-09-07  & 171   16:16-16:46     & A:16:40-17:15  & NPCH, EDGE\\
     &             & 195   16:25-16:35     & B:16:40-17:20  & $\alpha(B)=18^\circ$ \\
     &$29.68^\circ$& 284   16:26           &  faint in A    & $\beta(B)=26^\circ$\\
     &             & 304   16:16-16:46     &                & \bfseries{Lambda-Hel}              \\
\hline
32   & 2007-09-09  & 171   13:08-14:16     & A:13:35-15:00  & NPCH, L/I\\
     &             & 195   13:05-14:15     & B:13:35-15:05  & $\alpha(A)=10^\circ$ \\
     &$30.04^\circ$& 284   13:26-13:46     &                & $\beta(A)=16^\circ$\\
     &             & 304   13:36-13:56     &                & \bfseries{ET-Hel}              \\
\hline
33   & 2007-09-11  & 171   12:18-12:56     & A:12:55-14:05  & NPCH, EDGE \\       
     &             & 195   12:25-12:55     & B:12:45-14:35  & $\alpha(A)=26^\circ$ \\
     &$30.41^\circ$& 284   12:26-12:46     &                & $\beta(A)=42^\circ$\\
     &             & 304   12:16-12:56     &                & \bfseries{?}               \\
\hline
34   & 2007-09-15  & 171   14:46-14:56     & A:15:05-15:45  & SPCH, EDGE/L\\
     &             & 195   14:55           & B:15:05-16:05  & $\alpha(A)=164^\circ$ \\
     &$31.17^\circ$& 284   14:46           &                & $\beta(A)=153^\circ$\\
     &             & 304   14:46-14:56     &                & \bfseries{Lambda}               \\
\hline 
35   & 2007-09-23  & 171   11:26-11:50     & A:12:05-13:15  & NPCH, I/L   \\
     &             & 195   11:36-11:55     & B:11:45-13:15  & $\alpha(A)=346^\circ$ \\
     &$32.59^\circ$& 284   11:26-11:46     &                & $\beta(A)=331^\circ$\\
     &             & 304   11:26-11:56     &                & \bfseries{Lambda}               \\
\hline
36   & 2007-09-28  & 171   11:06-11:16     & A:11:25-12:25  & NPCH, EDGE/L\\ 
     &             & 195   11:05-11:15     & B:11:20-12:50  & $\alpha(A)=344^\circ$ \\
     &$33.46^\circ$& 284   11:06-11:26     &                & $\beta(A)=332^\circ$\\
     &             & 304   11:06-11:16     &                & \bfseries{ET-Hel}              \\
\hline
37   & 2007-10-01  & 171   23:26:23:36     & A:23:30-00:30  & NPCH, L  \\  
     &             & 195   23:25           & B:             & $\alpha(A)=359^\circ$ \\ 
     &$34.06^\circ$& 284   23:26           &                & $\beta(A)=352^\circ$\\
     &             & 304   23:26-23:36     &                & \bfseries{?}              \\
\hline
38   & 2007-10-05  & 171   08:06-09:06     & A:08:40-10:00  & NPCH, I   \\
     &             & 195   08:05-09:05     & B:08:35-09:55  & $\alpha(A)=354^\circ$ \\ 
     &$34.62^\circ$& 284   08:06-08:46     &                & $\beta(A)=349^\circ$\\
     &             & 304   08:06-09:06     &                & \bfseries{?}              \\
\hline       
39   & 2007-10-06  & 171   06:06-06:26     & A:06:05-07:35  & SPCH, EDGE\\
     &             & 195   06:15-06:25     & B:06:05-08:05  & $\alpha(A)=207^\circ$ \\ 
     &$34.77^\circ$& 284   06:06-06:26     &                & $\beta(A)=224^\circ$\\
     &             & 304   06:06-06:26     &                & \bfseries{?}              \\
\hline
40   & 2007-10-06  & 171   20:21-20:38     & A:20:45-21:55  & NPCH, I  \\   
     &             & 195   20:15-20:35     & B:20:35-21:15  & $\alpha(A)=354^\circ$ \\
     &$34.86^\circ$& 284   20:26           &                & $\beta(A)=348^\circ$\\
     &             & 304   20:26-20:46     &                & \bfseries{ET-Hel}              \\ 
\hline
41   & 2007-10-12  & 171   23:51-00:16     & A:00:05-01:00  & SPCH, I  \\
     &             & 195   23:55-00:25     & B:             & $\alpha(A)=182^\circ$ \\
     &$35.84^\circ$& 284                   &                & $\beta(A)=184^\circ$\\ 
     &             & 304   23:56-00:56     &                & \bfseries{Micro-CME-Hel}              \\  
\hline
42   & 2007-10-18  & 171   15:34-15:59?    & A:             & SPCH, L  \\ 
     &             & 195   15:45-15:55     & B:             & $\alpha(A)=178^\circ$ \\
     &$36.71^\circ$& 284   15:46           &(diff.movie(A):16:05)& $\beta(A)=180^\circ$\\
     &             & 304   15:36-16:06     & very faint     & \bfseries{ET}              \\ 
\hline
43   & 2007-10-21  & 171   22:16-22:56     & A:22:40-23:20  & SPCH, L\\ 
     &             & 195   22:15-22:25     & B:22:45-23:30  & $\alpha(A)=184^\circ$ \\
     &$37.19^\circ$& 284   22:06-22:46     &                & $\beta(A)=186^\circ$\\        
     &             & 304   22:26-22:56     &                & \bfseries{ET-Hel}              \\
\hline
44   & 2007-11-01  & 171   00:38-00:46     & A:00:55-02:05  & SPCH, L   \\              
     &             & 195   00:45           & B:01:05-02:05  & $\alpha(A)=169^\circ$ \\
     &$38.58^\circ$& 284   00:46-01:26     &                & $\beta(A)=172^\circ$\\
     &             & 304   00:46-01:26     &     faint      & \bfseries{ET}              \\
\hline
45   & 2007-11-01  & 171   02:21-02:46     & A:02:55-03:45  & SPCH, L   \\
     &             & 195   02:25-02:55     & B:             & $\alpha(A)=168^\circ$ \\ 
     &$38.59^\circ$& 284   02:26           &                & $\beta(A)=163^\circ$\\
     &             & 304   02:26-02:46     & not seen in B  & \bfseries{ET}              \\
\hline
46   & 2007-11-03  & 171   04:06-04:39     & A:04:45-05:35  & SPCH, L   \\
     &             & 195   04:05-04:45     & B:04:45-05:45  & $\alpha(A)=191^\circ$ \\
     &$38.86^\circ$& 284   04:06-04:46     &                & $\beta(A)=196^\circ$\\
     &             & 304   04:06-04:56     &                & \bfseries{Lambda-Hel}              \\
\hline
47   & 2007-11-04  & 171   01:56-02:56     & A:02:55-  ?    & NPCH, L   \\
     &             & 195   01:55-02:55     & B:             & $\alpha(A)=355^\circ$ \\
     &$38.97^\circ$& 284   02:06           & very faint     & $\beta(A)=354^\circ$\\
     &             & 304   02:06.03:06     &                & \bfseries{ET-Hel}              \\
\hline
48   & 2007-11-06  & 171   00:28-00:46     & A:00:45-02:15  & SPCH, L   \\     
     &             & 195   00:35-00:45     & B:00:45-02:05  & $\alpha(A)=176^\circ$ \\
     &$39.22^\circ$& 284                   &                &  $\beta(A)=177^\circ$\\  
     &             & 304   00:36-00:56     &                & \bfseries{ET}              \\
\hline
49   & 2007-11-09  & 171   02:53-03:36     & A:03:25-04:35  & NPCH, L    \\ 
     &             & 195                   & B:03:15-04:35  & $\alpha(A)=348^\circ$ \\  
     &$39.60^\circ$& 284   02:46           &                & $\beta(A)=332^\circ$\\ 
     &             & 304   02:56-03:46     &                & \bfseries{Micro-CME}              \\
\hline
50   & 2007-11-09  & 171   21:36-21:53     & A:21:55-23:15  & SPCH, EDGE/L\\ 
     &             & 195   21:36-22:16     & B:22:05-22:35  & $\alpha(A)=151^\circ$ \\
     &$39.69^\circ$& 284                   &                & $\beta(A)=147^\circ$\\
     &             & 304   21:46-22:16     &                & \bfseries{Lambda}              \\ 
\hline
51   & 2007-11-17  & 171   17:08-17:23     & A:17:25-19:15  & NPCH, I  \\
     &             & 195   17:15-17:45     & B:17:25-18:55  & $\alpha(A)=6^\circ$ \\
     &$40.57^\circ$& 284   17:26           &                & $\beta(A)=7^\circ$\\
     &             & 304   17:16-17:26     &                & \bfseries{Lambda-Hel}              \\ 
\hline
52   & 2007-11-19  & 171   09:53-10:21     & A:10:45-11:55  & NPCH, EDGE \\
     &             & 195   09:55-10:25     & B:10:35-12:05  & $\alpha(A)=328^\circ$ \\
     &$40.75^\circ$& 284   10:06           &                & $\beta(A)=327^\circ$\\
     &             & 304   09:56-10:26     &                & \bfseries{?}              \\
\hline
53   & 2007-11-21  & 171   21:28-22:26     & A:22:15-23:35  & NPCH, EDGE/L\\ 
     &             & 195   21:35-22:26     & B:             & $\alpha(A)=18^\circ$ \\
     &$41.01^\circ$& 284   21:26-21:46     &                & $\beta(A)=34^\circ$\\
     &             & 304   21:16-22:26     &                & \bfseries{?-Hel}              \\
\hline
54   & 2007-11-26  & 171   22:53-23:16     & A:23:15-23:55? & SPCH, L  \\   
     &             & 195   22:45-23:25     & B:23:25-23:55? & $\alpha(A)=191^\circ$ \\ 
     &$41.51^\circ$& 284                   &                & $\beta(A)=196^\circ$\\ 
     &             & 304   22:56-23:26     &                & \bfseries{?-Hel}              \\ 
\hline
55   & 2007-12-01  & 171                   & A:14:55-16:55  & NPCH, I  \\
     &             & 195   14:45-15:25     & B:14:45-17:05  & $\alpha(A)=10^\circ$ \\ 
     &$41.93^\circ$& 284   15:06           &                & $\beta(A)=18^\circ$\\ 
     &             & 304   14:46-15:06     &                & \bfseries{ET}              \\
\hline
56   & 2007-12-02  & 171                   & A:13:45-15:15  & NPCH, I   \\
     &             & 195   13:15-13:35     & B:13:35-15:15  & $\alpha(A)=10^\circ$ \\
     &$42.01^\circ$& 284                   &                & $\beta(A)=18^\circ$\\
     &             & 304   13:06-13:26     &                & \bfseries{ET}              \\
\hline
57   & 2007-12-03  & 171                   & A:01:55-02:35  & SPCH, L   \\
     &             & 195   01:25           & B:01:55-03:35  & $\alpha(B)=178^\circ$ \\
     &$42.06^\circ$& 284                   &                & $\beta(B)=167^\circ$\\
     &             & 304   01:26-01:46     &                & \bfseries{?}              \\
\hline
58   & 2007-12-03  & 171   03:51-04:26     & A:04:05-05:15  & SPCH, L   \\
     &             & 195   03:55           & B:04:15-05:15  & $\alpha(A)=191^\circ$ \\
     &$42.07^\circ$& 284                   &                & $\beta(A)=197^\circ$\\
     &             & 304   03:46-04:16     &                & \bfseries{?-Hel}              \\
\hline
59   & 2007-12-03  & 171   04:46-05:16     & A:05:05-06:35? & NPCH, L   \\
     &             & 195   04:45-05:25     & B:05:05-06:05? & $\alpha(A)=10^\circ$ \\
     &$42.07^\circ$& 284   04:46-05:26     &                & $\beta(A)=16^\circ$\\
     &             & 304   04:46-05:26     &                & \bfseries{?-Hel}              \\ 
\hline
60   & 2007-12-04  & 171   01:06-01:21     & A:not visible  & SPCH, L   \\      
     &             & 195   00:55-01:05     & B:01:15-02:00  & $\alpha(B)=180^\circ$ \\ 
     &$42.14^\circ$& 284                   & faint          & $\beta(B)=181^\circ$\\
     &             & 304   01:06-01:36     & (diff.movie(B):01:25)& \bfseries{ET}              \\
\hline
61   & 2007-12-05  & 171   10:53-11:26     & A:11:25-12:45  & NPCH, I   \\
     &             & 195   11:05-11:25     & B:11:15-12:55  & $\alpha(A)=7^\circ$ \\ 
     &$42.26^\circ$& 284                   &                & $\beta(A)=9^\circ$\\
     &             & 304   10:56-11:26     &                & \bfseries{?-Hel}              \\
\hline
62   & 2007-12-06  & 171   15:31-16:09     & A:             & NPCH, I   \\
     &             & 195   15:05-15:35     & B:             & $\alpha(B)=352^\circ$ \\
     &$42.36^\circ$& 284                   &(diff.movie(B):16:05)& $\beta(B)=348^\circ$\\ 
     &             & 304   15:16-16:16     &                & \bfseries{ET-Hel}              \\
\hline              
63   & 2007-12-06  & 171   19:21-19:36     & A:             & NPCH, L  \\
     &             & 195   19:45           & B:             & $\alpha(B)=355^\circ$ \\
     &$42.37^\circ$& 284   19:26           &(diff.movie(B):19:45-20:05)&$\beta(B)=351^\circ$  \\
     &             & 304   19:26-19:46     &                & \bfseries{?}              \\
\hline
64   & 2007-12-09  & 171   07:56           & A:08:15-10:15  & SPCH, L  \\ 
     &             & 195   07:55-08:25     & B:08:25-09:45  & $\alpha(A)=174^\circ$ \\
     &$42.57^\circ$& 284                   &                & $\beta(A)=172^\circ$\\
     &             & 304   07:56-08:26     &                & \bfseries{ET}              \\ 
\hline
65   & 2007-12-11  & 171   19:56-20:13     & A:20:05-21:00? & NPCH, L   \\ 
     &             & 195   19:35-19:45     & B:20:05-21:05  & $\alpha(A)=0^\circ$ \\ 
     &$42.76^\circ$& 284   20:06           &                & $\beta(A)=0^\circ$\\
     &             & 304   18:46-20:16     &                & \bfseries{?}              \\
\hline
66   & 2007-12-12  & 171   21:56-22:38     & A:             & SPCH, L   \\
     &             & 195   21:55           & B:             & $\alpha(A)=178^\circ$ \\
     &$42.84^\circ$& 284   22:06           & not visible    & $\beta(A)=177^\circ$\\ 
     &             & 304   21:56-22:46     & (diff.movie(A):22:25)& \bfseries{ET-Hel}      \\
\hline
67   & 2007-12-12  & 171   22:21-22:46?    & A:             & NPCH, EDGE\\
     &             & 195   22:25-22:46     & B:             & $\alpha(A)=31^\circ$ \\ 
     &$42.84^\circ$& 284   22:26           & (diff.movie(A):23:05)  &$\beta(A)=56^\circ$\\ 
     &             & 304   22:26-22:56     & faint signature& \bfseries{Lambda}             \\ 
\hline
68   & 2007-12-17  & 171   04:11-04:28     & A:             & SPCH, EDGE \\
     &             & 195   04:15-04:45     & B:             & $\alpha(A)=201^\circ$ \\ 
     &$43.14^\circ$& 284   04:26           & (diff.movie(A):04:25-& $\beta(A)=216^\circ$\\
     &             & 304   04:16-04:46     & 05:05)& \bfseries{ET}\\
\hline
69   & 2007-12-21  & 171   04:48-05:03     & A:             & NPCH, I  \\
     &             & 195   04:55-05:15     & B:05:15-06:25  & $\alpha(B)=353^\circ$ \\
     &$43.40^\circ$& 284                   &                & $\beta(B)=351^\circ$\\
     &             & 304   04:46-05:16     &                & \bfseries{ET-Hel}              \\
\hline
70   & 2007-12-22  & 171   03:03-03:38     & A:             & NPCH, L  \\
     &             & 195   03:05-03:35     & B:03:20-04:50  & $\alpha(B)=351^\circ$ \\ 
     &$43.46^\circ$& 284   03:26           &                & $\beta(B)=347^\circ$\\
     &             & 304   03:06-03:46     &                & \bfseries{ET-Hel}              \\
\hline
71   & 2008-01-09  & 171   07:36-07:57     & A:08:05-08:55  & NPCH, L  \\ 
     &             & 195   07:45-08:00     & B:07:55-09:00  & $\alpha(A)=351^\circ$ \\ 
     &$44.41^\circ$& 284   07:36-07:56     &                & $\beta(A)=344^\circ$\\
     &             & 304   07:41-08:06     &                & \bfseries{ET-Hel}              \\
\hline
72   & 2008-01-20  & 171   00:16-00:31     & A:00:25-01:35  & NPCH, L/I \\
     &             & 195   00:15-00:30     & B:00:20-01:10  & $\alpha(B)=1^\circ$ \\ 
     &$44.83^\circ$& 284   00:16           &                & $\beta(B)=9^\circ$\\
     &             & 304   00:17-00:39     & very faint in A& \bfseries{ET}              \\
\hline
73   & 2008-01-20  & 171   16:26-17:11     & A:             & NPCH, EDGE \\
     &             & 195   16:30-17:10     & B:             & $\alpha(B)=340^\circ$ \\
     &$44.86^\circ$& 284   16:26-17:11     & (diff.movie(B):17:10) &$\beta(B)=321^\circ$\\
     &             & 304   16:28-17:12     &                & \bfseries{Micro-CME-Hel?}              \\
\hline
74   & 2008-02-02  & 171   13:18-13:41     & A:13:35-14:15  & NPCH, L  \\ 
     &             & 195   13:25-13:35     & B:13:35-14:05  & $\alpha(B)=352^\circ$ \\  
     &$45.27^\circ$& 284   13:26           &                & $\beta(B)=349^\circ$\\ 
     &             & 304   13:16-13:56     & very faint in A& \bfseries{Lambda-Hel}              \\ 
\hline
75   & 2008-02-08  & 171   08:33-08:43     & A:08:45-09:35  & NPCH, L   \\
     &             & 195   08:35:08:45     & B:08:40-10:00  & $\alpha(A)=351^\circ$ \\
     &$45.45^\circ$& 284                   &                & $\beta(A)=342^\circ$\\
     &             & 304   08:36-08:56     &                & \bfseries{?-Hel}              \\ 
\hline
76   & 2008-02-17  & 171   03:36-03:56     & A:04:05-05:35  & SPCH, I  \\
     &             & 195   03:35-03:55     & B:             & $\alpha(A)=159^\circ$  \\
     &$45.70^\circ$& 284                   & not recognizable&    $\beta(A)=164^\circ$\\
     &             & 304   03:36-04:06     & in B           &  \bfseries{ET}             \\
\hline
77   & 2008-03-02  & 171   07:41-08:26     & A:07:55-08:55  & NPCH, L   \\
     &             & 195   07:45-08:25     & B:07:55-08:55  & $\alpha(A)=14^\circ$ \\
     &$46.16^\circ$& 284   07:46-08:06     &                & $\beta(A)=31^\circ$\\
     &             & 304   07:36-07:46     &                & \bfseries{?}              \\
\hline
78   & 2008-03-27  & 171   06:23-07:06     & A:06:45-07:55  & NPCH, I/EDGE\\
     &             & 195   06:25-07:05     & B:06:35-07:45  & $\alpha(A)=2^\circ$ \\
     &$47.26^\circ$& 284   06:26-06:46     &                & $\beta(A)=357^\circ$\\
     &             & 304   06:26-07:06     &                & \bfseries{ET-Hel}              \\
\hline
79   & 2008-04-02  & 171   08:31-08:46     & A:08:55-10:05  & SPCH, L \\
     &             & 195   08:35-09:15     & B:08:55-10:15  & $\alpha(A)=185^\circ$ \\
     &$47.62^\circ$& 284                   &                & $\beta(A)=188^\circ$ \\
     &             & 304   08:36-08:56     &                & \bfseries{ET-Hel}              \\ 
\hline
\hline
\end{longtable}


%
\begin{acks}
It is a pleasure to thank all the STEREO staff, without which this work would not have been possible. G. N. acknowledges support from an Erasmus grant during his stay in Goettingen. V.B acknowledges the support of the project Stereo/Corona by the German Bundesministerium f\"{u}r Bildung und Forschung through the deutsche Zentrum f\"{u}r Luft-und Raumfahrt e.V. (DLR, German Space Agency) as a collaborative effort with the Max-Planck-Institut f\"{u}r Sonnensystemforschung (MPS) under grant 50 0C 0904. Stereo/Corona is a science and hardware contribution to the optical image package SECCHI, developed for the NASA STEREO mission launched in 2006. The SECCHI data used here were produced by an international consortium of the Naval Research Laboratory (USA), Lockheed Martin Solar and Astrophysics Lab (USA), NASA Goddard Space Flight Center (USA), Rutherford Appleton Laboratory (UK), University of Birmingham (UK), Max-Planck-Institut for Solar System Research (Germany), Centre Spatiale de Li\'ege (Belgium), Institut d’Optique Th\'eorique et Appliqu\'ee (France), and Institut d’Astrophysique Spatiale (France). G. Z. was supported in part by the Italian INAF and by the Italian Space Agency, contract ASI n. I/015/07/0 ``Esplorazione del Sistema Solare''. We would like to thank the referee for his/her careful and constructive criticism which has allowed us to substantially improve the paper.

\end{acks}

%
%
%

\end{article} 
\end{document}